\documentclass[superscriptaddress,twocolumn,secnumarabic,
nobibnotes,aps,prd,showpacs,nofootinbib]{revtex4}
\usepackage{graphicx}
\usepackage{epsf}
\usepackage{bm}
\usepackage{amsmath}
\usepackage{amsfonts}
\usepackage{amssymb}
\usepackage{epstopdf}
\usepackage{natbib}
\usepackage{color}
\usepackage{color}
\providecommand{\U}[1]{\protect\rule{.1in}{.1in}}
\newcommand{\be}{\begin{equation}}
\newcommand{\ee}{\end{equation}}

\newcommand{\mincir}{\raise
-3.truept\hbox{\rlap{\hbox{$\sim$}}\raise4.truept\hbox{$<$}\ }}
\newcommand{\magcir}{\raise
-3.truept\hbox{\rlap{\hbox{$\sim$}}\raise4.truept\hbox{$>$}\ }}

\begin{document}

\title{Field theoretic interpretations of interacting dark energy scenarios and recent observations}

\author{Supriya Pan}
\email{supriya.maths@presiuniv.ac.in}
\affiliation{Department of Mathematics, Presidency University, 86/1 College Street, Kolkata 700073, India.}

\author{German S. Sharov}
\email{Sharov.GS@tversu.ru}
\affiliation{Tver State University, 170002, Sadovyj per. 35, Tver, Russia}

\author{Weiqiang Yang}
\email{d11102004@163.com}
\affiliation{Department of Physics, Liaoning Normal University, Dalian, 116029, P. R. China.}

\begin{abstract}
Cosmological models describing the non-gravitational interaction between dark matter and dark energy  are based on some phenomenological choices of the interaction rates between dark matter and dark energy. There is no such guiding rule to select such rates of interaction. {\it In the present work we show that various phenomenological models of the interaction rates might have a strong field theoretical ground.} We explicitly derive several well known interaction functions between dark matter and dark energy under some special conditions and finally constrain them using the latest cosmic microwave background observations from final Planck legacy release together with baryon acoustic oscillations distance measurements. Our analyses report that one of the interacting functions is able to alleviate the $H_0$ tension. We also perform a Bayesian evidence analyses for all the models with reference to the $\Lambda$CDM model. From the Bayesian evidence analyses, although the reference scenario is preferred over the interacting scenarios, however, we found that two interacting models are close to the reference $\Lambda$CDM model.

\end{abstract}

\maketitle


\section{Introduction}

Observational evidences from various astronomical sources suggest that a non-zero interaction in the dark sectors, i.e., between dark matter (DM) and dark energy (DE) is allowed
\cite{Bolotin:2013jpa,Wang:2016lxa}, and consequently, a mild deviation
from the non-interacting $\Lambda$-cosmology is expected.
Although within $1\sigma$ confidence level one may recover
$\Lambda$CDM model, but however, the null-interaction is not yet confirmed. The question arises why should we consider the interaction between DM and DE? The answer could be given in different ways. Since the nature and evolution of DM and DE are not known to us then there is no justification to avoid the possibility of mutual interaction between these dark sectors. In fact,
the interaction in the dark sector is a promising approach which solves the coincidence
problem \cite{Amendola:1999er,TocchiniValentini:2001ty,Amendola:2003eq,delCampo:2008sr,delCampo:2008jx} and it was motivated to solve the cosmological constant problem \cite{Wetterich-ide1}.
Investigations by several investigators in the last couple of years explored some more interesting properties of the interacting DE models \cite{Cai:2004dk,Barrow:2006hia,He:2008tn,Valiviita:2008iv,Majerotto:2009np,Valiviita:2009nu,Clemson:2011an,Nunes:2014qoa,Salvatelli:2014zta,Yang:2014gza,Yang:2014hea,Pan:2013rha,Pan:2014afa,Pan:2012ki,Nunes:2016dlj,Kumar:2016zpg,Yang:2016evp,vandeBruck:2016hpz,Pan:2016ngu,Sharov:2017iue,Yang:2017yme,Guo:2017hea,Shahalam:2017fqt,Yang:2017ccc,Yang:2017zjs,Pan:2017ent,Li:2017usw,Amendola:2017xhl,Begue:2017lcw,Yang:2018pej,Yang:2018ubt,Yang:2018xlt,Li:2018ydj,vonMarttens:2018iav,Yang:2018qec,Martinelli:2019dau,Asghari:2019qld,Paliathanasis:2019hbi,Pan:2019jqh,Yang:2019bpr,Zhang:2019zxv,Yang:2019vni,Yang:2019uzo,Savastano:2019zpr,Pan:2019gop,DiValentino:2019jae,Yang:2019uog,vonMarttens:2019ixw,Papagiannopoulos:2019kar}. It has been shown that the interaction between DM and DE can solve
the tension on the present Hubble constant value, $H_0$, appearing from its local and
global measurements \cite{Kumar:2017dnp,DiValentino:2017iww,Yang:2018euj,Yang:2018uae,Kumar:2019wfs,DiValentino:2019ffd} and also the tension in the amplitude of the matter power spectrum, $\sigma_8$, by different observations \cite{Kumar:2019wfs,vandeBruck:2017idm,Barros:2018efl}.
However, although the models in such theory are phenomenologically
motivated, but nevertheless, from the particle physics point of view, the interaction
between DM and DE is a natural phenomenon because any two fields (DM field and DE field) can interact with each other. In the last several
years, several people have studied the DM and DE dynamics with
different choices for the interaction function relating the energy densities
of the dark sectors. Depending on the choice of the function, the interaction
becomes linear or non-linear in the energy densities of the dark sectors.

Mathematically and physically as well, we have no rigid theoretical bounds for the mentioned interaction functions. If the
universe contains $n$ matter components with the energy momentum tensors $T^i_{\mu\nu}$,
$i=1,\dots,n$, such that either all or some of the energy components interact with each other, then the energy conservation condition
$$\nabla^\mu\sum_{i=1}^n T^i_{\mu\nu}=0$$
is fulfilled only for all matter, but not for every $i$-th component.
 So we can add and subtract any interaction function $Q_\nu\equiv Q_\nu^{ij}$ for
 $i$-th and $j$-th components:
\begin{equation}
 \nabla^\mu T^i_{\mu\nu}=Q_\nu,\qquad \nabla^\mu T^j_{\mu\nu}=-Q_\nu.
\label{Q1}
\end{equation}

The investigators mentioned earlier (see again \cite{Bolotin:2013jpa,Wang:2016lxa,Amendola:1999er,TocchiniValentini:2001ty,Amendola:2003eq,delCampo:2008sr,delCampo:2008jx,Wetterich-ide1,Cai:2004dk,Barrow:2006hia,He:2008tn,Nunes:2014qoa,Salvatelli:2014zta,Yang:2014gza,Yang:2014hea,Pan:2013rha,Pan:2014afa,Pan:2012ki,Nunes:2016dlj,Kumar:2016zpg,Yang:2016evp,vandeBruck:2016hpz,Pan:2016ngu,Sharov:2017iue,Yang:2017yme,Guo:2017hea,Shahalam:2017fqt,Yang:2017ccc,Yang:2017zjs,Pan:2017ent,Li:2017usw,Amendola:2017xhl,Begue:2017lcw,Yang:2018pej,Yang:2018ubt,Yang:2018xlt,Li:2018ydj,vonMarttens:2018iav,Yang:2018qec,Martinelli:2019dau,Asghari:2019qld,Paliathanasis:2019hbi,Pan:2019jqh,Yang:2019bpr,Zhang:2019zxv})
worked with different phenomenological variants of the interaction function $Q$. In
particular, in the interactive models \cite{Amendola:1999er,Wetterich-ide1,Carroll:1998zi,Billyard:2000bh,Nunes:2000ka} the DE
component was described as a scalar field $\phi$, that takes part in the
phenomenologically constructed interaction with the standard cold DM (an ideal fluid with zero pressure).

In this paper we suggest
a variant of motivation to consider the interaction term $Q$
in different linear and non-linear forms. Our approach includes
a symmetric description of both DM and DE as
two scalar fields $\phi_1$ and $\phi_2$, where  they may interact via their common potential
$V(\phi_1,\phi_2)$. It is  widely known that scalar fields can simulate cosmological
evolution (see reviews \cite{Copeland:2006wr, Bamba:2012cp}). Models with two scalar
fields were suggested and studied in
Refs.~\cite{Elizalde:2004mq, Nojiri:2005sx, Nojiri:2005pu, Capozziello:2005tf, Feng:2004ad}, however,
the authors' interest was not concentrated on the possibilities to describe interaction of dark components.

In this approach we suppose, that the interaction function $Q$ can be deduced from the
(more fundamental) potential $V(\phi_1,\phi_2)$.
 The connection between $V$ and  $Q$ is rather complicated in general, in particular, the linear dependence of $Q$ on densities (the linear interaction) is not the most
obvious result of this approach. In any case we have a degree of freedom on a certain
level: when we choose the potential $V$, or the interaction term $Q$.

We organize the work as follows. In section \ref{sec-2}, we give the details
of the mathematical formulation of an interacting universe
 and describe, how different forms of the interaction function $Q$ can be deduced
 from the common potential of scalar fields.
In section \ref{sec-data_and_results}, we list the observational data to analyse the
models and consequently present the results of various observational analyses.
Finally, we summarize the main findings of the work in section  \ref{sec-summary}.

\section{Interacting dark energy: A field theoretic description}
\label{sec-2}

We consider a cosmological scenario where two heavy dark fluids in the universe,
namely, the DM and DE non-gravitationally interact with other. The other components, namely the baryons and radiation do not take part in the interaction. To describe such interacting universe, as usual we assume the homogeneous and isotropic Friedmann-Lema\^{i}tre-Robertson-Walker line element given by
\begin{eqnarray}
d{\rm s}^2 = -dt^2 + a^2 (t) \left[\frac{dr^2}{1-\kappa r^2} + r^2 \left(d \theta^2 + \sin^2
\theta d \phi^2\right)  \right]. \label{FLRW}
\end{eqnarray}
Here, $a(t)$ is the scale factor of this FLRW universe and $\kappa$ is
the curvature sign of the universe. The curvature
sign  may describe three different geometries of
the universe, namely, flat ($\kappa =0$), open ($\kappa = -1$) and closed ($\kappa=1$).
Since most of the observational estimations prefer a flat
geometry of the universe, see for instance \cite{Ade:2015xua,Aghanim:2018eyx}, henceforth, we fix the spatial flatness of the universe in this work. Thus, in such a prescribed geometric structure of the universe one can write down the Einstein's field equations as
\begin{eqnarray}
&& H ^2 = \frac{8 \pi G}{3} \sum _{i} \rho_{i},\label{Fried}\\
&& \dot{H} = - 4 \pi G \sum _{i} (p_{i} +\rho_{i}),\label{dotH}
\end{eqnarray}
where an overhead dot in any quantity denotes
its cosmic time differentiation; $H \equiv \dot{a}/a$
is the Hubble rate of this FLRW universe; ($\rho_i$, $p_i$)
respectively refer to the energy density and pressure of the
$i$-th fluid. Precisely, $\rho_r$, $\rho_b$, $\rho_c$, $\rho_x$ are
respectively the energy densities of radiation, baryons, DM and a DE
fluid. Similarly, $p_r$, $p_b$, $p_c$ and $p_x$ are respectively the pressure components of radiation, baryons, DM and DE. Since radiation and baryons do not take part in the interaction, thus they follow the standard
evolution equations while the conservation equations of the interacting DM and DE
follow,

\begin{eqnarray}
\dot{\rho}_c + 3 H ( \rho_c+p_c) &=& -Q, \label{con01}\\
\dot{\rho}_x + 3 H(\rho_x+p_x)  &=& Q. \label{con02}
\end{eqnarray}
Below we shall assume that DM is the cold DM (abbreviated as CDM) or pressure-less and thus, $p_c =0$. The interactive term $Q$ in Eqs.~(\ref{con01}), (\ref{con02}) is usually factorized by the Hubble rate $H$ and can depend on the densities $\rho_c$, $\rho_x$, pressures
$p_c$, $p_x$, other parameters
\cite{Wetterich-ide1,Carroll:1998zi,Amendola:1999er,Billyard:2000bh,Nunes:2000ka}
(see also the classification in Ref.~\cite{Bolotin:2013jpa}). We can choose a function
$Q(H,\rho_c,\rho_x,\dots)$ in many variants: different possibilities of this choice and
a more fundamental approach to deduce $Q$  may be illustrated in the following scheme.

We generalize the traditional scalar field simulation of DE
\cite{Wetterich-ide1, Carroll:1998zi,Amendola:1999er,Billyard:2000bh,Nunes:2000ka,Copeland:2006wr, Bamba:2012cp}
and suppose that both DM and DE fluids are described correspondingly
as two real scalar fields $\phi_1$ and $\phi_2$. Their interaction is naturally managed
by a common potential $V(\phi_1,\phi_2)$ in the action
\cite{Elizalde:2004mq, Nojiri:2005sx, Nojiri:2005pu, Capozziello:2005tf}
\begin{eqnarray}
S=\int d^4x\sqrt{-g}\Bigg[\frac R{16\pi G}-\frac{\epsilon_1}2
(\nabla\phi_1)^2 -\frac{\epsilon_2}2 (\nabla\phi_2)^2 \nonumber\\- V(\phi_1,\phi_2)
\Bigg]+S^m.\label{action}
\end{eqnarray}
Here, $R=g^{\mu\nu}R_{\mu\nu}$ is the Ricci scalar, the factors $\epsilon_j=\pm1$
determine quintessential or phantom nature of a field,
$(\nabla\phi_j)^2=g^{\mu\nu}\partial_\mu\phi_j\partial_\nu\phi_j$, the term $S^m$
describes the remaining matter (baryons, radiation).

The action (\ref{action}) is symmetric with respect to DM $\phi_1$ and DE $\phi_2$. This form is convenient to generate an interaction of these fluids,
however, it does not coincide with the widely used approach
\cite{Wetterich-ide1,Carroll:1998zi,Amendola:1999er,Billyard:2000bh,Nunes:2000ka,Copeland:2006wr, Bamba:2012cp},
where a scalar field $\phi$ describes only the DE. For DM one can
find some form of a scalar field description in mimetic models
\cite{Chamseddine:2013kea,Chamseddine:2014vna} (also see \cite{Myrzakulov:2015qaa,Cognola:2016gjy,Sebastiani:2016ras,Vagnozzi:2017ilo,deHaro:2018sqw}), but in the action (\ref{action}) the
field  $\phi_1$ is not connected with conformal degrees of freedom of any auxiliary
metric.

If we vary the action (\ref{action}) over $g^{\mu\nu}$, $\phi_1$ and $\phi_2$, we deduce
the dynamical equations
\begin{eqnarray}
R_{\mu\nu}-\frac R2 g_{\mu\nu}= 8\pi G \Bigg[\sum_{j=1}^2
\epsilon_j\Big(\partial_\mu\phi_j\partial_\nu\phi_j-\frac12(\nabla\phi_j)^2g_{\mu\nu}\Big) \nonumber\\-Vg_{\mu\nu}+T^m_{\mu\nu}\Bigg], \label{eqR}\\
\nabla^\mu\nabla_\mu\phi_j=\epsilon_j\frac{\partial V}{\partial\phi_j},\qquad
j=1,2.\label{eqphi}
\end{eqnarray}
The covariant divergence of the equation (\ref{eqR}) leads to the energy conservation
equation $\nabla^\mu T^m_{\mu\nu}=0$ for baryons and radiation, because the terms with
$\phi_j$ vanish as the consequence of  Eqs.~(\ref{eqphi}).

For the FLRW universe (\ref{FLRW}) in its flat case $\kappa =0$ the equations
(\ref{eqR}) may be reduced to the form (\ref{Fried}), (\ref{dotH}), but with the
following new content of the total density and pressure:
\begin{eqnarray}
\rho_{tot} = \frac{\epsilon_1}2 {\dot\phi_1}^2+\frac{\epsilon_2}2
{\dot\phi_2}^2+V(\phi_1,\phi_2) + \rho_b +\rho_r,\nonumber\\
 p_{tot} = \frac{\epsilon_1}2
{\dot\phi_1}^2+\frac{\epsilon_2}2 {\dot\phi_2}^2-V(\phi_1,\phi_2) + p_b + p_r.
 \label{rhotot}
\end{eqnarray}

The scalar field equations (\ref{eqphi}) for the FLRW universe (\ref{FLRW}) take the
similar form
\begin{eqnarray}
\ddot\phi_1+3H\dot\phi_1&=&-\epsilon_1\frac{\partial V}{\partial\phi_1}, \label{eqphi1}\\
\ddot\phi_2+3H\dot\phi_2&=&-\epsilon_2\frac{\partial V}{\partial\phi_2}, \label{eqphi2}
\end{eqnarray}
but they describe interaction of the fluids $\phi_1$ and $\phi_2$, if the potential
$V(\phi_1,\phi_2)$ is not equal to a sum $V_1(\phi_1)+V_2(\phi_2)$.

This interaction between $\phi_1$ and $\phi_2$ can be rewritten and represented in the
form (\ref{con01}), (\ref{con02}), if we divide the common potential $V(\phi_1,\phi_2)$
into two parts (introducing an additional degree of freedom with this division)
\begin{equation}
V(\phi_1,\phi_2)=V_1(\phi_1,\phi_2)+V_2(\phi_1,\phi_2)\label{V1V2}
\end{equation}
and determine the densities and pressures of the dark components:
\begin{eqnarray}
 \rho_1=\frac{\epsilon_1}2 {\dot\phi_1}^2+V_1,\quad p_1=\frac{\epsilon_1}2 {\dot\phi_1}^2-V_1,\nonumber\\
 \rho_2=\frac{\epsilon_2}2 {\dot\phi_2}^2+V_2,\quad
  p_2=\frac{\epsilon_2}2 {\dot\phi_2}^2-V_2~.\label{rho12}
\end{eqnarray}

In these notations the dynamical  equations  (\ref{eqphi1}), (\ref{eqphi2}) for 2 scalar
fields will take the form (\ref{con01}), (\ref{con02}):
\begin{equation}
 \begin{array}{l}
\dot{\rho}_1 + 3 H ( \rho_1+p_1) =-Q, \\
\dot{\rho}_2 + 3 H(\rho_2+p_2)  = Q,\rule{0pt}{1.2em}
 \end{array}
\label{conQV12}
\end{equation}
with the interacting term
\begin{equation}
Q=\dot\phi_1\frac{\partial V_2}{\partial\phi_1}- \dot\phi_2\frac{\partial
V_1}{\partial\phi_2}.\label{QV12}
\end{equation}

Obviously, this interaction term $Q$ equals zero in the case of non-interacting
decomposing potential \cite{Feng:2004ad}
\begin{equation}
V(\phi_1,\phi_2)=V_1(\phi_1)+V_2(\phi_2).
 \label{V1V2nint}
\end{equation}

In the general case (\ref{V1V2}) we have the  mentioned degree of freedom, when $V$ is
divided into $V_1$ and $V_2$. However, this degree of freedom is a form of gauge
transformations and does not change the model behavior: if we redefine $V_1$ to $\tilde
V_1=V_1+\delta V(\phi_1,\phi_2)$ (and $V_2\;\to\;\tilde V_2=V_2-\delta V$), we will
obtain the correspondent redefinition of $\rho_1$ and $Q$, in particular,
$Q\;\to\;\tilde Q=Q-\frac d{dt}\delta V$. But the dynamical equations (\ref{Fried}),
(\ref{eqphi1}), (\ref{eqphi2}) and observable manifestations will remain just the same.

The most surprising point in the considered model (\ref{action}) is its description of
the cold DM (an ideal fluid with zero pressure) as the scalar field  $\phi_1$.
However, it is possible, if we, naturally, fix the sign $\epsilon_1=1$ and require zero
value for the pressure $p_1$ (\ref{rho12}):
\begin{equation}
\epsilon_1=1,\qquad p_1\equiv\frac12 {\dot\phi_1}^2-V_1(\phi_1,\phi_2)=0.
 \label{p10}
\end{equation}
 In particular, this model without the DE field ($\phi_2=0$, $V_2=0$) under the
 condition (\ref{p10}) recovers the Friedmann solution $a=(t/t_0)^{2/3}$ with the
 following exponential potential:
 \begin{eqnarray}
&& V_1(\phi_1)\equiv V(\phi_1)=A\exp\big(-2\sqrt{6\pi G}\,\phi_1\big),\nonumber\\
&&\phi_1=\frac{\log t}{\sqrt{6\pi G}} +\mbox{const}.
\end{eqnarray}
The $\Lambda$CDM model will be reproduced, if for the non-interacting case
(\ref{V1V2nint}) we impose the conditions (\ref{p10}),
 $\phi_2=0$, $V_2={}$const, and obtain
\begin{eqnarray}
&&\phi_2= 0, \qquad V_2=\frac\Lambda{8\pi G}=\mbox{const},
\end{eqnarray}
\begin{eqnarray}
V_1(\phi_1)= \frac{\Lambda}{16\pi G}\sinh^2\big[\sqrt{6\pi G}\,(\phi_1-\phi_0)\big]\nonumber\\
= \frac{\Lambda}{16\pi G}\sinh^{-2}\big[\sqrt{3\Lambda}\,(t-t_0)\big].
\end{eqnarray}
Here, $\phi_0$, $t_0$ are constants of integration.

Another solution for two interacting scalar fields was obtained in
Ref.~\cite{Nojiri:2005sx}, it describes the Big Rip singularity (at $t=t_s$) with the
Hubble parameter, fields
 \begin{eqnarray}
&&H =\frac\theta3\bigg(\frac1t+ \frac1{t-t_s} \bigg),\;
 \phi_1=\phi_0\log\frac t{t_0},\nonumber\\
&&\phi_2=\phi_0\log\frac{t_s-t}{t_0},\qquad\epsilon_2=-1
  \label{solNOT}
\end{eqnarray}
and the potential
 \begin{eqnarray}
V(\phi_1,\phi_2)=\frac{\phi_0^2}{2t_0^2}\Bigg[(\theta-1)\,e^{-2\phi_1/\phi_0}+(\theta+1)\,e^{-2\phi_2/\phi_0} \nonumber\\ +2\theta e^{-(\phi_1+\phi_2)/\phi_0}\Bigg].
  \label{VNOT}
\end{eqnarray}
Here, $\phi_0$, $t_0$ are constants, and $\theta=12\pi G\phi_0^2$. The solutions
(\ref{solNOT}), (\ref{VNOT}) do not satisfy the condition (\ref{p10}).

If we divide the term with $ e^{-(\phi_1+\phi_2)/\phi_0}$ in the  potential (\ref{VNOT})
symmetrically between $V_1$ and $V_2$ in Eq.~(\ref{V1V2}), the interaction function $Q$
(\ref{QV12}) will be
$$
Q=-\frac{\phi_0^2\theta}{2t_0^3}\big(e^{-\phi_1/\phi_0}+e^{-\phi_2/\phi_0}\big)\,
e^{-(\phi_1+\phi_2)/\phi_0}.
$$
It may be expressed via the Hubble parameter (\ref{solNOT}) and the densities
(\ref{rho12}) $\rho_j=\frac12\phi_0^2\theta
t_0^{-2}\big(e^{-\phi_1/\phi_0}+e^{-\phi_2/\phi_0}\big)\, e^{-\phi_j/\phi_0}$ as
follows:
 \begin{equation}
Q = 3\xi H  \frac{\rho_1 \rho_2}{\rho_1 + \rho_2}.
  \label{Q20}
\end{equation}
Here $\xi=-1/\theta$. The interaction function (\ref{Q20}) may be deduced in another way \cite{Nojiri:2005sx}, if we require for the system (\ref{con01}), (\ref{con02}), that the ratios $r=\rho_c/\rho_x$,  $w_c = p_c/\rho_c$ and $w_x = p_x/\rho_x$ are constants. These conditions for Eqs.~(\ref{con01}), (\ref{con02}) lead to the equality
$$
\frac d{dt}\log r=-\frac Q{\rho_c}-\frac Q{\rho_x}+3 H (w_x-w_c)=0,
$$
that is equivalent to Eq.~(\ref{Q20}).

 Another variant of an interacting model may be described by a potential of the type (\ref{VNOT}),
however, we impose the condition (\ref{p10}) and choose $V_1(\phi_1)$ depending only on
$\phi_1$ as follows:
 \begin{eqnarray}
V(\phi_1,\phi_2)=V_1(\phi_1)+V_2(\phi_1,\phi_2)=\frac{\phi_0^2}{2t_1^2}\,e^{-2\phi_1/\phi_0}\nonumber\\+ A_2t_1^{\gamma_1}t_2^{\gamma_2} e^{\gamma_1\phi_1/\phi_0+\gamma_2\phi_2/\psi_0}.
  \label{Vexp}
  \end{eqnarray}
Here, $\phi_0$, $\psi_0$, $A_2$, $t_j$, $\gamma_j$ are constants and
 $$\gamma_1+\gamma_2=-2.$$
 We consider the solution
  \begin{equation}
 \phi_1=\phi_0\log\frac t{t_1},\;
 \phi_2=\psi_0\log\frac{t}{t_2}, \;
  H =\frac{h_0}t,\, h_0=\mbox{const},
  \label{sol2}
\end{equation}
 unlike Eq.~(\ref{solNOT}) it has no  future singularity,  but under the condition
(\ref{p10}), that means $p_1=0$, we obtain
 $$
 \rho_1=\dot\phi_1^2=2V_1=\frac{\phi_0^2}{t^2}, \qquad
 V_2=\frac{A_2}{t^2}. $$

  We use equations (\ref{Fried}),  (\ref{eqphi1}), (\ref{eqphi2}) to express the constants
 in Eqs.~(\ref{Vexp}), (\ref{sol2}) via dimensionless parameters $\gamma_1$ and $h_0$:
  \begin{eqnarray}
 \phi_0^2=\frac{\gamma_1h_0(3h_0-1)}{8\pi G(2-3h_0+\gamma_1/2)},\nonumber\\
 \epsilon_2\psi_0^2=\frac{(2+\gamma_1)h_0(2-3h_0)}{8\pi G(2-3h_0+\gamma_1/2)},
 \nonumber\\
  A_2=\frac{h_0(3h_0-1)(2-3h_0)}{8\pi G(2-3h_0+\gamma_1/2)}.
  \label{expr2}
\end{eqnarray}

Note that in the case $h_0=2/3$ interaction vanishes and we have the potential
(\ref{V1V2nint}), that means, $V=V_1(\phi_1)+V_2(\phi_2)$ (or $V_2=0$). But for  $h_0\ne\frac23$
 the interaction term (\ref{QV12}) $Q=\dot\phi_1\frac{\partial V_2}{\partial\phi_1}$
is nonzero and may be presented in the following forms:
\begin{eqnarray}
Q = 3H \xi_1\rho_1, &\quad&\xi_1=\frac{2-3h_0}{3h_0},   \label{Q21}\\
Q = 3H \xi_2\rho_2, &\quad &\xi_2=\frac{\gamma_1(3h_0-1)}{3h_0(3h_0+\gamma_1/2)},
\label{Q22}
 \end{eqnarray}
or in the form (\ref{Q20}) with
$\xi=\big(2-3h_0+\frac{\gamma_1}2\big)\big/\big(3h_0+\frac{\gamma_1}2\big)$.

Here, the `DE component' density $\rho_2$ is proportional to $\rho_1$, but
pressure $p_2$ remains nonzero
  \begin{equation}
 \rho_2=\frac{h_0(2-3h_0)(3h_0+\gamma_1/2)}{8\pi G(2-3h_0+\gamma_1/2)\,t^2},\qquad
 p_2=\frac{h_0(2-3h_0)}{8\pi Gt^2}.
   \label{rp2} \end{equation}

 The densities $\rho_2$ and $\rho_1=\phi_0^2/t^2$ with $\phi_0$ from Eq.~(\ref{expr2}) are positive in the following three physical cases:
 \begin{eqnarray}
 \mbox{(a)}&\;& h_0>\frac23,\; -6h_0< \gamma_1 < 0 \; \Rightarrow \; Q<0;\nonumber\\
 \mbox{(b)}&\;& \frac13<h_0<\frac23,\; \gamma_1 > 0 \;\Rightarrow\; Q>0;\nonumber\\
 \mbox{(c)}&\;& 0<h_0<\frac13,\; \max\{-6h_0,6h_0-4\}<\gamma_1<0\; \Rightarrow\; Q>0.  \nonumber
 \end{eqnarray}
 In the cases (b) and (c) we deal only with a quintessential DE
 ($\epsilon_2=1$) and obtain $Q>0$, but in the variant (a)  we have $Q<0$
 and can construct this fluid with both signs of $\epsilon_2$.

We consider DM as cold (that means pressure-less) under the condition (\ref{p10}) with  $p_c =0$ ($p_c \equiv p_1$), and DE as vacuum ($p_x  = - \rho_x$). So the equations (\ref{con01}) and (\ref{con02}) for $\rho_c \equiv\rho_1$ and
$\rho_x \equiv\rho_2$ take the form

\begin{eqnarray}
\dot{\rho}_c + 3 H  \rho_c &=& -Q, \label{cons1}\\
\dot{\rho}_x   &=& Q,\label{cons2}
\end{eqnarray}
where $Q$ is the interaction function that has been already mentioned earlier. The sign of the interaction rate has a physical meaning. For positive values of the interaction rate, that means for $Q >0$, the transfer of energy and/or momentum takes place from pressureless DM to DE while its opposite sign that means, $Q <0$ refers to the opposite case, i.e.,  energy flow takes place from DE to pressureless DM. 
Now, for any arbitrary given interaction function, $Q$, using the above conservation equations (\ref{cons1}), (\ref{cons2}) together with the Hubble equation (\ref{Fried}), one can solve the evolution of ($\rho_c$, $\rho_x$) either analytically or numerically.  Usually, for any arbitrary interaction function, the background evolution of ($\rho_c$, $\rho_x$) cannot be analytically found. For some specific interaction models, it is possible to impose some analytic structure on the background evolution of the dark sectors' energy densities.

\section{Interaction models and their perturbations}
\label{sec-perturbations}

As already shown  in section \ref{sec-2}, the field theoretic approach returns some very well known interaction models in which the interaction function is either linearly related to their individual energy density, namely, $Q \propto \rho_c$, $Q \propto \rho_x$ or it could  have a nonlinear structure  involving the energy densities of the dark sectors  as follows $Q \propto \rho_c \rho_x (\rho_c + \rho_x)^{-1}$. So, one can clearly justify that the linear combination of the energy densities of the dark components could be  another feasible interaction model emerging from this context. Thus, in the present article, we shall consider four distinct interaction functions shown in Table \ref{tab:models} and explore the corresponding cosmological scenarios.  Before that we discuss  the linear perturbations for any interaction model which are very essential to understand the overall cosmological picture driven by an interaction function.              

\begin{table*}
\begin{center}
\renewcommand{\arraystretch}{1.4}
\begin{tabular}{|c@{\hspace{1 cm}}|@{\hspace{1 cm}} c|@{\hspace{1 cm}} c|}
\hline
\textbf{Model No.}                    & \textbf{Expression for $Q$} & \textbf{Corresponding Cosmic Scenario} \\
\hline\hline

Model I  & $3 H \xi \rho_c$ & IVS0 \\
Model II & $3 H \xi \rho_x$ & IVS1 \\
Model III & $3 H \xi (\rho_c+\rho_x)$ & IVS2\\
Model IV & $3 H \xi \left(\frac{\rho_c \rho_x}{\rho_c+\rho_x}\right)$ & IVS3\\

\hline
\end{tabular}
\end{center}
\caption{We show the interaction models that we shall study in this work. In the table we have clearly labeled the interacting scenario corresponding to some specific interaction function. Here, IVSi ($i =0, 1, 2, 3$) means the Interacting Vacuum Scenario for the $i$-th interaction function. }
\label{tab:models}
\end{table*}
To start with the perturbations equations for any interaction models we consider the perturbed FLRW metric given by  \cite{Mukhanov:1990me,Ma:1995ey,Malik:2008im}
\begin{eqnarray}\label{perturbed-metric}
ds^{2}=-(1+2\phi )dt^{2}+2  a \partial _{i}B dt dx^{i}\notag\\+
a^{2}\Bigl((1-2\psi )\delta _{ij}+2\partial _{i}\partial _{j}E\Bigr)dx^{i}dx^{j},
\end{eqnarray}
where $\phi $, $B$, $\psi $, $E$, are
the gauge-dependent scalar perturbation quantities. Here, we work with the synchronous gauge, that means, we set $\phi = B = 0$, $\psi =\eta $, and $k^{2}E=-h/2-3\eta $, where $k$ is the Fourier mode and $h$, $\eta$ are the metric perturbations. Now, let us discuss how the perturbations equations look like when the above metric (\ref{perturbed-metric}) is considered. To start with the perturbations equations, let us now be more explicit this time only by focusing on the interaction between CDM and the vacuum energy, and hence, let us recast the equation (\ref{Q1}) as 

\begin{eqnarray}
\label{DE_DM_1}
\nabla_{\mu}T_{i}^{\mu \nu } =Q_{i}^{\nu}\,, \quad \sum\limits_{\mathrm{i}}{%
Q_{i}^{\nu }}=0~,
\end{eqnarray}
where the index $i$ takes the values $\{c, x\}$ in which  as already mentioned    
$i = c$ stands the CDM component and $i=x$ stands for the vacuum energy component. We assume that the four vector $Q_{i}^{\nu}$, quantifying the interaction between these dark fluids, is given by 
\begin{eqnarray}
Q_{i}^{\nu}=(Q_{i}+\delta Q_{i})u^{\nu}+a^{-1}(0,\partial^{\nu}f_{i}), 
\end{eqnarray}
where $u^{\nu}$ is the velocity four vector and $Q_i$ denotes the  background energy transfer. We note that here  $Q_i \equiv Q$.  Also, $f_i$ represents the momentum transfer potential. Now, for the perturbed FLRW metric (\ref{perturbed-metric}), one can  derive the perturbations equations.  The continuity equations for CDM and vacuum as shown in eqn. (\ref{DE_DM_1}) can be found to be 
\begin{eqnarray}
\label{delta_DM}
\delta \dot{\rho_{c}} + 3 H \delta \rho_{c} - 3 \rho_c \dot{\psi} + \rho_{c} \frac{k^2}{a^2} \Big( \theta_{c} + a^2 \dot{E} \Big) = - \delta Q, 
\end{eqnarray}
\begin{eqnarray}
\label{delta_DE}
\delta \dot{\rho_{x}} = \delta Q,
\end{eqnarray}
while the conservation equations for the momentum for both CDM and vacuum energy become, 

\begin{eqnarray}
\label{theta_DM}
\rho_c \dot{\theta_{c}} = -f - Q(\theta - \theta_c),
\end{eqnarray}
\begin{eqnarray}
\label{theta_DE}
-\delta_{x} = f + Q \theta,
\end{eqnarray}
where let us note that $f = f_c =  f_x$ and $f$ is the momentum transfer. With the similar notations noted above, the symbols $f_c$, $f_x$ naturally associate to  the momentum transfer of CDM and vacuum energy.   
Combining now the equations (\ref{delta_DM}), (\ref{delta_DE}), and  (\ref{theta_DM}), (\ref{theta_DE}), one may eliminate $\delta Q$ and $f$ leading to the  following equations,

\begin{eqnarray}
\label{delta_DM_2}
\delta \dot{\rho_{c}} + 3 H \delta \rho_c - 3 \rho_c \dot{\psi} + \frac{k^2}{a^2} \Big(\theta_c + a^2 \dot{E} \Big) = - \dot{\delta_x}~,
\end{eqnarray}
\begin{eqnarray}
\label{theta_DM_2}
\rho_c \dot{\theta_{c}} = \delta_x + Q \theta_c.
\end{eqnarray}
now we assume the similar approach treated in \cite{Wang:2014xca}, that means we consider an energy flow parallel to the 4-velocity of CDM given by $Q^{\mu}_c = - Q^{\mu} u^{\mu}_c$. For this case, CDM follows geodesics \cite{Wang:2014xca} which consequently means that the vacuum energy perturbations will vanish in the CDM co-moving frame from eq. (\ref{theta_DM_2}). Therefore, from the residual gauge freedom in the synchronous gauge, one obtains that $\theta_c = 0$ and $\delta_x = 0$. As a result, in the co-moving synchronous gauge, the density perturbation equation for the CDM component takes,

\begin{eqnarray}
\label{delta_DM_3}
\dot{\delta_c} = - \frac{\dot{h}}{2} + \frac{Q}{\rho_c} \delta_c.
\end{eqnarray}

We also refer to Ref. \cite{Corasaniti:2008kx} for more details in this direction when some general interacting scenarios are considered. Thus, having the evolution equations of the dark sectors' energy densities at the level of background (see the discussions at the end of section \ref{sec-2}) and perturbations, it is possible to proceed to examine the interacting scenarios with the use of latest observational data. The next section is  devoted for this purpose.

\section{Observational data, statistical methodology and the results}
\label{sec-data_and_results}

We describe, in this section, the observational data and the methodology for the
statistical analyses of the models. We use the latest cosmic microwave background
observations from  Planck 2018 \cite{Aghanim:2018oex,Aghanim:2019ame} and baryon
acoustic oscillations distance measurements
\cite{Beutler:2011hx,Ross:2014qpa,Gil-Marin:2015nqa}. To constrain the interacting
scenarios, we use the \texttt{CosmoMC} package \cite{Lewis:2002ah,Lewis:1999bs}, a
 Markov chain Monte Carlo code used to extract the observational constraints. This code supports the
Planck 2018 likelihood and it has a valid convergence diagnostic by Gelman-Rubin \cite{Gelman-Rubin}. Here, we have modified this publicly available \texttt{CosmoMC} package with the present models, that means for the interaction functions. 
Since in all the interacting scenarios that we
consider in this work, vacuum interacts with cold DM, thus, the dimension of the
parameter space is seven with the following parameters:

\begin{eqnarray}
\mathcal{P} \equiv \Bigl\{\Omega_bh^2, \Omega_{c}h^2, 100\theta_{MC}, \tau, n_s, log[10^{10}A_s], \xi \Bigr\},\nonumber
\end{eqnarray}
where the first six parameters in  $\mathcal{P}$ refer to the six parameters of the $\Lambda$CDM model and the parameter $\xi$ is the coupling parameter, mentioned earlier.
In Table \ref{tab:priors}, we show the flat priors on various free parameters of the interacting cosmic scenarios during the 
statistical analysis.

\begin{table}
\begin{center}
\renewcommand{\arraystretch}{1.4}
\begin{tabular}{|c@{\hspace{1 cm}}|@{\hspace{1 cm}} c|}
\hline
\textbf{Parameter}                    & \textbf{Prior}\\
\hline\hline
$\Omega_{b} h^2$             & $[0.005,0.1]$\\
$\Omega_{c} h^2$             & $[0.01,0.99]$\\
$\tau$                       & $[0.01,0.8]$\\
$n_s$                        & $[0.5, 1.5]$\\
$\log[10^{10}A_{s}]$         & $[2.4,4]$\\
$100\theta_{MC}$             & $[0.5,10]$\\
$\xi$                        & $[-1, 1]$ \\
\hline
\end{tabular}
\end{center}
\caption{Flat priors on various free parameters of the interacting scenarios have been shown. }
\label{tab:priors}
\end{table}

\subsection{IVS0}

\begin{figure}
\includegraphics[width=0.48\textwidth]{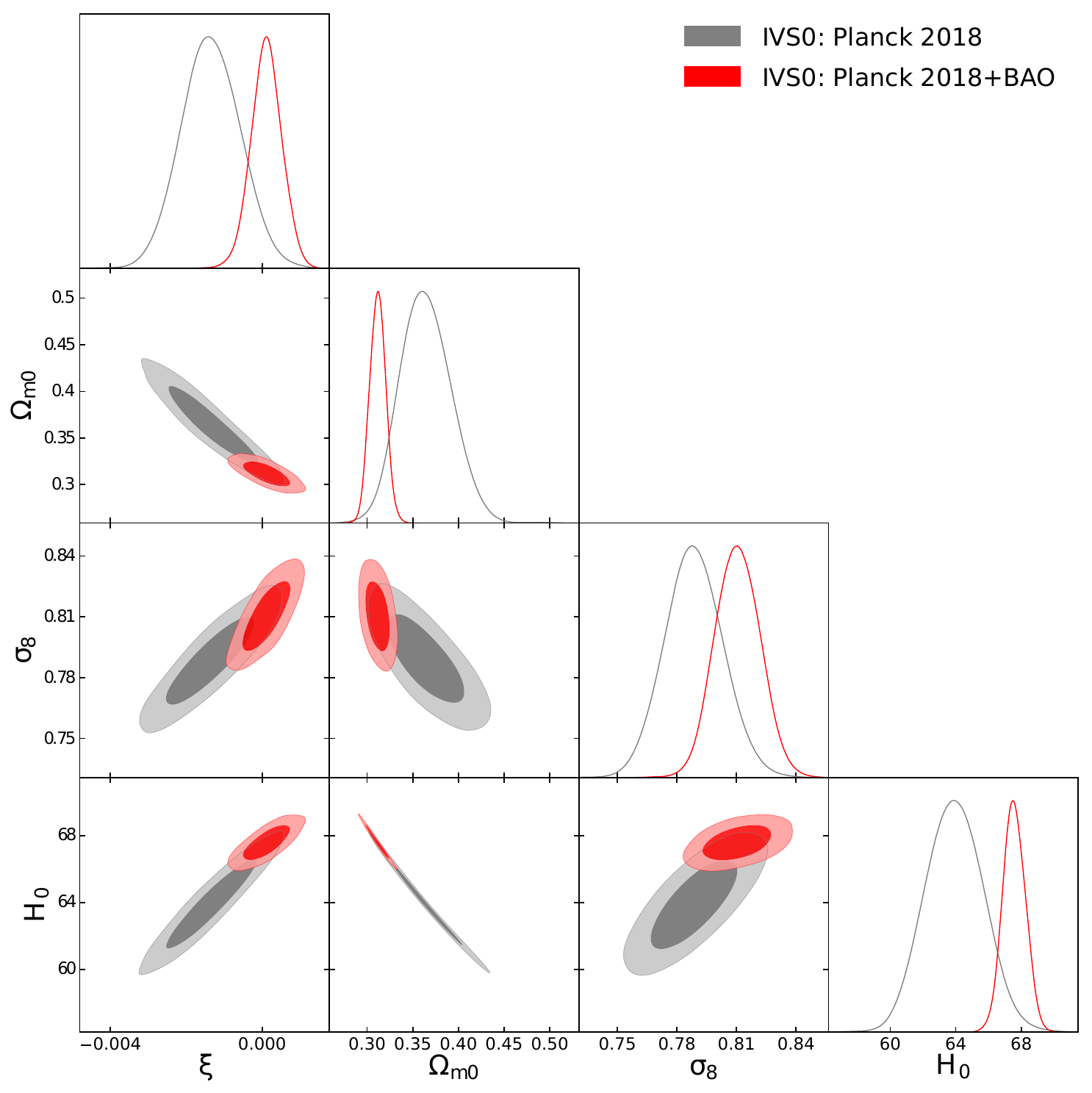}
\caption{One dimensional marginalized posterior distributions of some selective parameters and two-dimensional joint contours of various combinations of the model parameters for IVS0 have been displayed. }
\label{fig:ivs0}
\end{figure}

We show the observational constraints on this interacting cosmic scenario in Table \ref{tab:IVS0} and in Fig. \ref{fig:ivs0} for Planck 2018 alone and Planck 2018+BAO data. We include BAO with
Planck 2018  in order to break the degeneracies between the parameters. From Planck 2018 data alone we find that
$\xi \neq 0$ is allowed at more than 68\% CL ($\xi = -0.0013_{-    0.00077}^{+    0.00077}$). So, a very mild interaction in the dark sector is signaled by Planck data alone, however, when BAO data are added to Planck CMB, $\xi$ becomes very small compared to its estimation from Planck CMB alone, but within 68\% CL, $\xi = 0$ is consistent ($\xi = 0.00011_{-    0.00040}^{+    0.00040}$).
The Hubble constant, $H_0$ assumes very lower value  ($H_0 = 63.93_{3.38}^{+ 3.44}$, 95\% CL, Planck 2018) compared to Planck's $\Lambda$CDM based estimation \cite{Aghanim:2018eyx} but with high error bars as one can see. When BAO are added to Planck 2018, $H_0$ goes up with reduced error bars ($H_0 = 67.56_{- 1.31}^{+ 1.35}$, 95\%, Planck 2018+BAO) and becomes consistent with Planck's $\Lambda$CDM based estimation  \cite{Aghanim:2018eyx}. So, as we can see that the tension on $H_0$ is not reconciled within this
 interacting scenario.

Regarding the estimation of $\Omega_{m0}$, we find a very strong anti correlation with $H_0$, and hence this parameter behaves
 accordingly with the increase or reduction of the Hubble constant. An interesting observation from Fig. \ref{fig:ivs0}, specifically from the  joint contour $(\Omega_{m0}, \sigma_8)$ is that, after the addition of BAO data to Planck 2018, the contour becomes vertical offering no correlation between them, while we note that for Planck 2018 data alone, the correlation between these two parameters are existing.

In summary, for this
 interaction model we find a very mild interaction
in the dark sector which is much consistent with the non-interaction cosmology.

\begingroup
\squeezetable
\begin{center}
\begin{table}
\begin{tabular}{ccccccccccccccc}
\hline\hline
Parameters & Planck 2018 & Planck 2018+BAO \\ \hline

$\Omega_c h^2$ & $    0.1254_{-    0.0031-    0.0060}^{+    0.0031+    0.0061}$ & $    0.1192_{-    0.0012-    0.0023}^{+    0.0012+    0.0023}$\\

$\Omega_b h^2$ & $    0.02239_{-    0.00016-    0.00032}^{+    0.00016+    0.00032}$ & $    0.02235_{-    0.00016-    0.00030}^{+    0.00016+    0.00032}$  \\

$100\theta_{MC}$ & $    1.04023_{-    0.00035-    0.00081}^{+    0.00042+    0.00071}$ & $    1.04077_{-    0.00030-    0.00059}^{+    0.00030+    0.00060}$ \\

$\tau$ & $    0.053_{-    0.0081-    0.015}^{+    0.0075+    0.016}$ & $    0.057_{-    0.0089-    0.016}^{+    0.0079+    0.017}$ \\

$n_s$ & $    0.9666_{-    0.0050-    0.0100}^{+    0.0050+    0.0096}$ & $    0.9742_{-    0.0037-    0.0072}^{+    0.0037+    0.0074}$  \\

${\rm{ln}}(10^{10} A_s)$ & $    3.053_{-    0.015-    0.031}^{+    0.015+    0.033}$ & $    3.058_{-    0.016-    0.033}^{+    0.017+    0.036}$   \\

$\xi$ & $   -0.0013_{-    0.00077-    0.0015}^{+    0.00077+    0.0015}$ &  $    0.00011_{-    0.00040-    0.00080}^{+    0.00040+    0.00079}$  \\

$\Omega_{m0}$ & $    0.364_{-    0.031-    0.051}^{+    0.027+    0.057}$ & $    0.312_{-    0.0091-    0.017}^{+    0.0086+    0.017}$  \\

$H_0$ & $   63.93_{-    1.79-    3.38}^{+    1.78+    3.44}$ & $   67.56_{-    0.73-    1.31}^{+    0.69+    1.35}$   \\
\hline\hline
\end{tabular}
\caption{Observational constraints at 68\% and 95\% CL on IVS0 corresponding to the interaction function $Q = 3 H \xi \rho_c$, using the CMB data from Planck 2018 and the data from BAO. Note that $H_0$ is in the units of Km \, s$^{-1}$ \, Mpc$^{-1}$. }
\label{tab:IVS0}
\end{table}
\end{center}
\endgroup

\subsection{IVS1}

\begin{figure}
\includegraphics[width=0.48\textwidth]{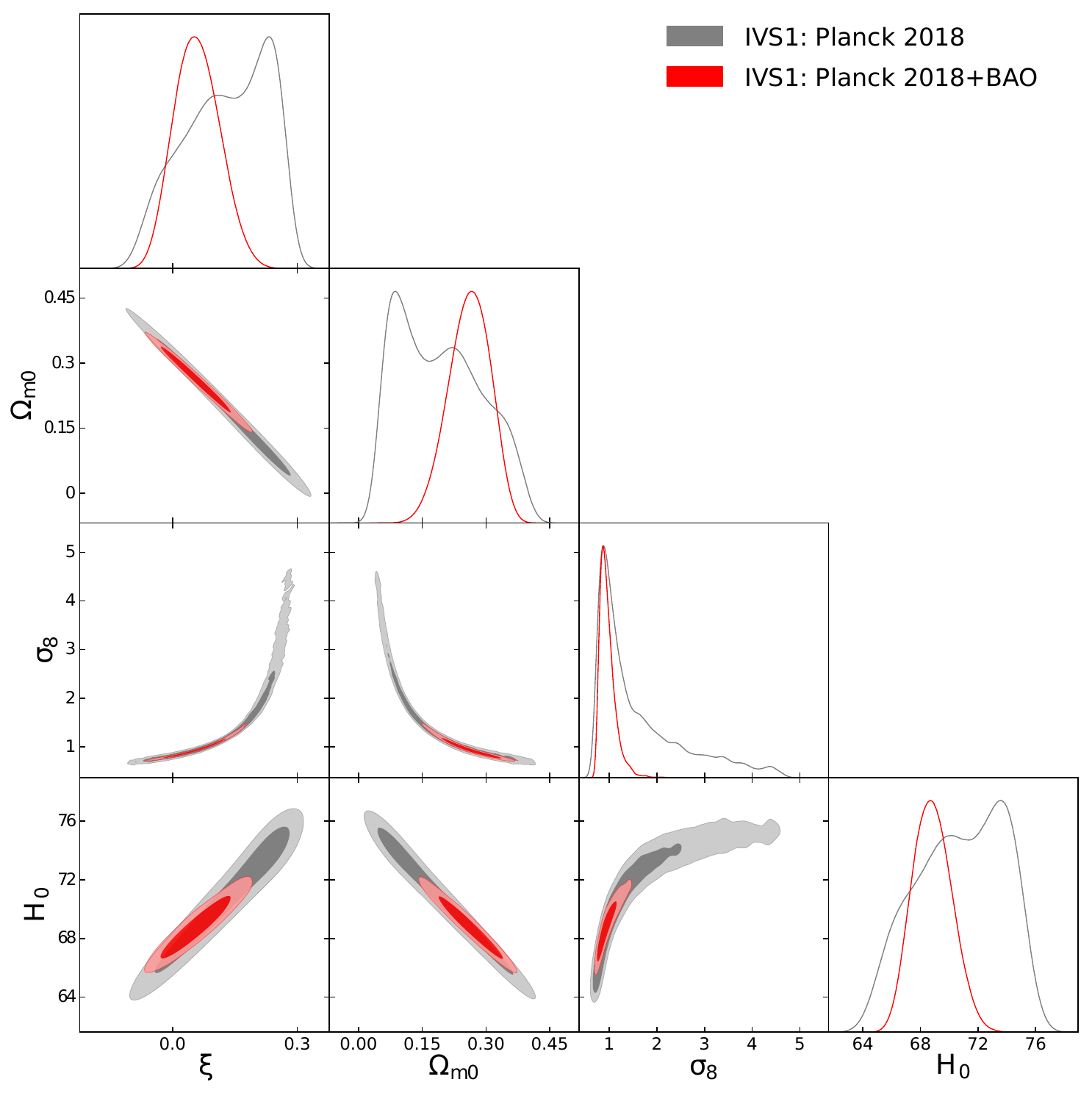}
\caption{One dimensional marginalized posterior distributions of some selective parameters and two-dimensional joint contours of various combinations of the model parameters for IVS1 have been displayed. }
\label{fig:ivs1}
\end{figure}

The summary of the observational constraints on this interaction model has been shown in Table \ref{tab:IVS1} and in Fig. \ref{fig:ivs1}.
One can clearly see that the estimations of the coupling parameter for both Planck 2018 and Planck 2018+BAO are  large compared to the previous interacting scenario IVS0. For Planck 2018, we see that at more than 68\% CL, $\xi \neq 0$ ($\xi = 0.132_{- 0.077}^{+    0.142}$, 68\% CL), but within 95\% CL, $\xi  =0$ is allowed. For Planck 2018+BAO, within 68\% CL, $\xi = 0$ is consistent. So, concerning the estimations of the coupling parameter,  we can safely conclude that a mild interaction is allowed within this interaction scenario.

Concerning the estimation of $H_0$, we find an interesting observation as follows. For Planck 2018, we find that $H_0$ takes a very high value with respect to the $\Lambda$CDM based Planck's estimation \cite{Aghanim:2018eyx} having in addition significantly high error bars that enables us it to reach its local estimation ($H_0 = 74.03 \pm 1.42$ Km\;$s^{-1}$\; $Mpc^{-1}$ at 68\% CL) \cite{Riess:2019cxk}, and thus, within 68\% CL, the tension on $H_0$ is clearly alleviated. When BAO data are added to Planck 2018, the mean value of $H_0$ is lowered with reduced error bars where $H_0 = 68.82_{-1.53}^{+    1.30}$ at 68\% CL. So, compared to the minimal $\Lambda$CDM cosmology, in this scenario the tension is less but not alleviated when we look at the measurement from \cite{Riess:2019cxk}.    

Thus, in summary, this interaction model has the ability to alleviate the $H_0$ tension offering a mild evidence of an interaction in the dark sector which is more pronounced for Planck 2018 alone.

\begingroup
\squeezetable
\begin{center}
\begin{table}
\begin{tabular}{cccccccccccccc}
\hline\hline
Parameters & Planck 2018 & Planck 2018+BAO \\ \hline

$\Omega_c h^2$ & $    0.0687_{-    0.0677-    0.0677}^{+    0.0244+    0.0647}$ & $    0.0996_{-    0.0156-    0.0383}^{+    0.0225+    0.0353}$  \\

$\Omega_b h^2$ & $    0.02230_{-    0.00015-    0.00029}^{+    0.00015+    0.00030}$ & $    0.02233_{-    0.00014-    0.00027}^{+    0.00014+    0.00028}$ \\

$100\theta_{MC}$ & $    1.04409_{-    0.00405-    0.00493}^{+    0.00258+    0.00548}$ & $    1.04188_{-    0.00134-    0.00207}^{+    0.00086+    0.00233}$  \\

$\tau$ & $    0.054_{-    0.0079-    0.015}^{+    0.0075+    0.015}$ & $    0.055_{-    0.0083-    0.016}^{+    0.0076+    0.016}$  \\

$n_s$ & $    0.9723_{-    0.0044-    0.0081}^{+    0.0043+    0.0083}$ & $    0.9734_{-    0.0040-    0.0078}^{+    0.0040+    0.0079}$  \\

${\rm{ln}}(10^{10} A_s)$ & $    3.055_{-    0.016-    0.030}^{+    0.015+    0.031}$ & $    3.057_{-    0.017-    0.032}^{+    0.016+    0.033}$  \\

$\xi$ & $    0.132_{-    0.077-    0.197}^{+    0.142+    0.169}$ & $    0.059_{-    0.061-    0.101}^{+    0.053+    0.110}$  \\

$\Omega_{m0}$ & $    0.191_{-    0.141-    0.166}^{+    0.075+    0.1901}$ & $    0.261_{-    0.046-    0.099}^{+    0.056+    0.095}$  \\

$H_0$ & $   70.84_{-    2.50-    5.94}^{+    4.26+    5.26}$ & $   68.82_{-    1.53-    2.64}^{+    1.30+    2.77}$  \\

\hline\hline
\end{tabular}
\caption{Observational constraints at 68\% and 95\% CL on IVS1 corresponding to the interaction function $Q = 3 H \xi \rho_x$, using the CMB data from Planck 2018 and the data from BAO. Note that $H_0$ is in the units of Km \, s$^{-1}$ \, Mpc$^{-1}$.  }
\label{tab:IVS1}
\end{table}
\end{center}
\endgroup

\subsection{IVS2}

\begin{figure}
\includegraphics[width=0.48\textwidth]{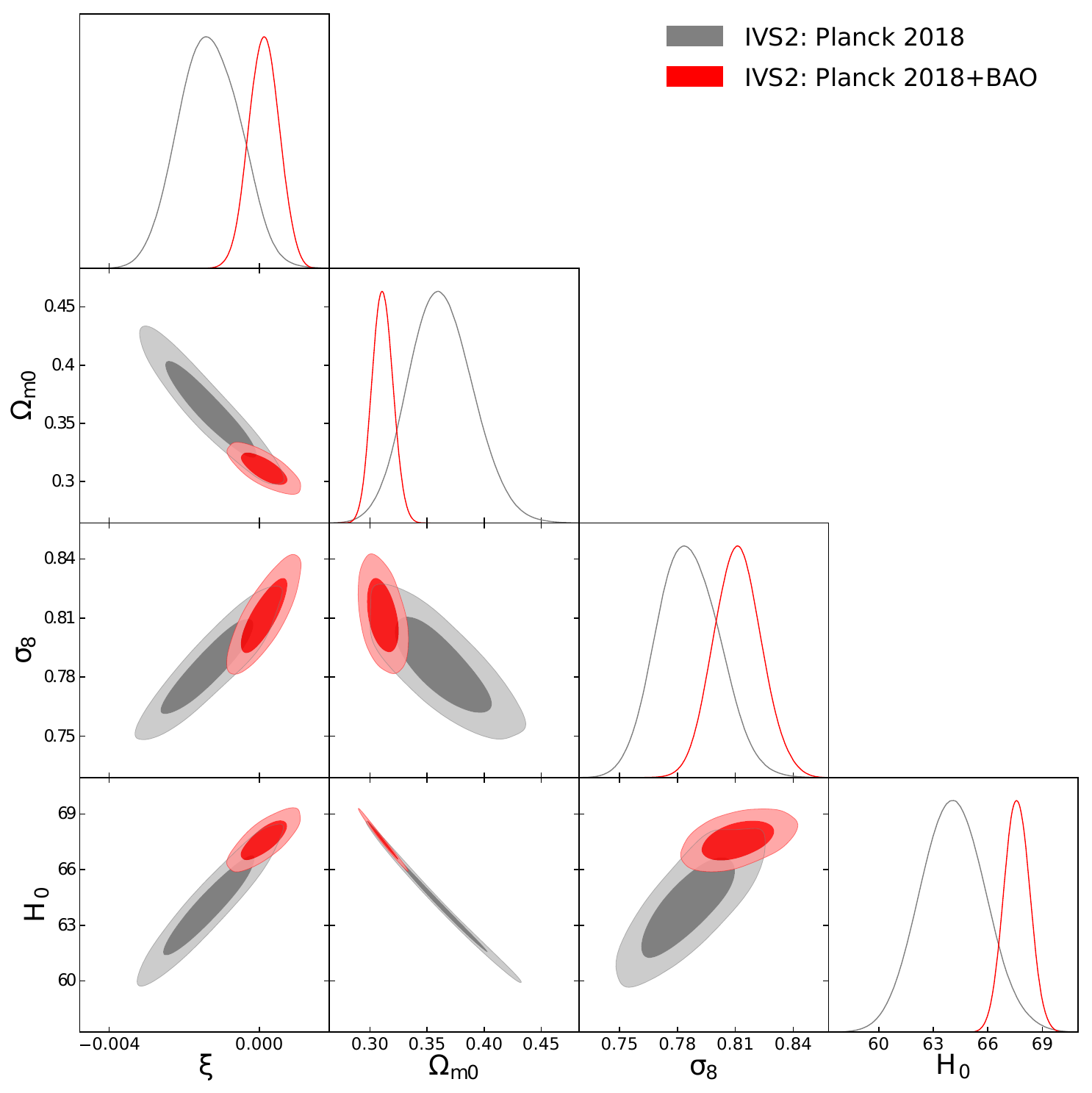}
\caption{One dimensional marginalized posterior distributions of some selective parameters and two-dimensional joint contours of various combinations of the model parameters for IVS2 have been displayed. }
\label{fig:ivs2}
\end{figure}

We show the observational constraints for this interacting scenario in Table \ref{tab:IVS2} and in Fig. \ref{fig:ivs2}. In a similar fashion we concentrate on the key
parameters $\xi$ and $H_0$ for this model. Note that the observational constraints for this model are almost similar to the IVS0 scenario. Thus, similar to IVS0 scenario, for this interaction model, an evidence of a non-null interaction at more than 68\% CL is favored ($\xi =  -0.0013_{- 0.00081}^{+    0.00085}$, 68\% CL, Planck 2018) for Planck 2018 data. While for Planck 2018+BAO, $\xi =0$ is consistent within
68\% CL.

Concerning the estimations of $H_0$ for both Planck 2018 and Planck 2018+BAO, we find that for Planck 2018, it takes lower values with high error bars unlike to what we find in $\Lambda$CDM based estimation \cite{Aghanim:2018eyx}. Thus, the tension on $H_0$ remains true for Planck 2018 data. However, when BAO data are added to Planck 2018, $H_0$ slightly goes up with reduced error bars, but effectively the $H_0$ tension is not alleviated.

\begingroup
\squeezetable
\begin{center}
\begin{table}
\begin{tabular}{cccccccccccc}
\hline\hline
Parameters & Planck 2018 & Planck 2018+BAO \\ \hline

$\Omega_c h^2$ & $    0.1253_{-    0.0032-    0.0065}^{+    0.0032+    0.0061}$ & $    0.1191_{-    0.0013-    0.0024}^{+    0.0012+    0.0024}$\\

$\Omega_b h^2$ & $    0.02243_{-    0.00019-    0.00036}^{+    0.00017+    0.00036}$ & $    0.02234_{-    0.00016-    0.00031}^{+    0.00016+    0.00032}$\\

$100\theta_{MC}$ & $    1.04023_{-    0.00037-    0.00075}^{+    0.00038+    0.00070}$ & $    1.04077_{-    0.00031-    0.00059}^{+    0.00030+    0.00058}$ \\

$\tau$ & $    0.052_{-    0.0073-    0.014}^{+    0.0072+    0.014}$ & $    0.057_{-    0.0085-    0.015}^{+    0.0074+    0.016}$ \\

$n_s$ & $    0.9678_{-    0.0051-    0.0097}^{+    0.0052+    0.0093}$ & $    0.9744_{-    0.0037-    0.0074}^{+    0.0036+    0.0073}$  \\

${\rm{ln}}(10^{10} A_s)$ & $    3.052_{-    0.015-    0.030}^{+    0.015+    0.029}$ & $    3.058_{-    0.017-    0.031}^{+    0.016+    0.033}$  \\

$\xi$ & $   -0.0013_{-    0.00081-    0.0015}^{+    0.00085+    0.0015}$ & $    0.00013_{-    0.00041-    0.00078}^{+    0.00040+    0.00078}$ \\

$\Omega_{m0}$ & $    0.362_{-    0.031-    0.053}^{+    0.027+    0.057}$ & $    0.311_{-    0.0092-    0.017}^{+    0.0090+    0.018}$ \\

$H_0$ & $   64.09_{-    1.81-    3.42}^{+    1.78+    3.58}$ & $   67.59_{-    0.68-    1.36}^{+    0.71+    1.38}$  \\
\hline\hline
\end{tabular}
\caption{Observational constraints at 68\% and 95\% CL on IVS2 corresponding to the interaction function $Q = 3 H \xi (\rho_c+\rho_x)$, using the CMB data from Planck 2018 and the data from BAO. Note that $H_0$ is in the units of Km \, s$^{-1}$ \, Mpc$^{-1}$. }
\label{tab:IVS2}
\end{table}
\end{center}
\endgroup

\subsection{IVS3}

\begin{figure}
\includegraphics[width=0.48\textwidth]{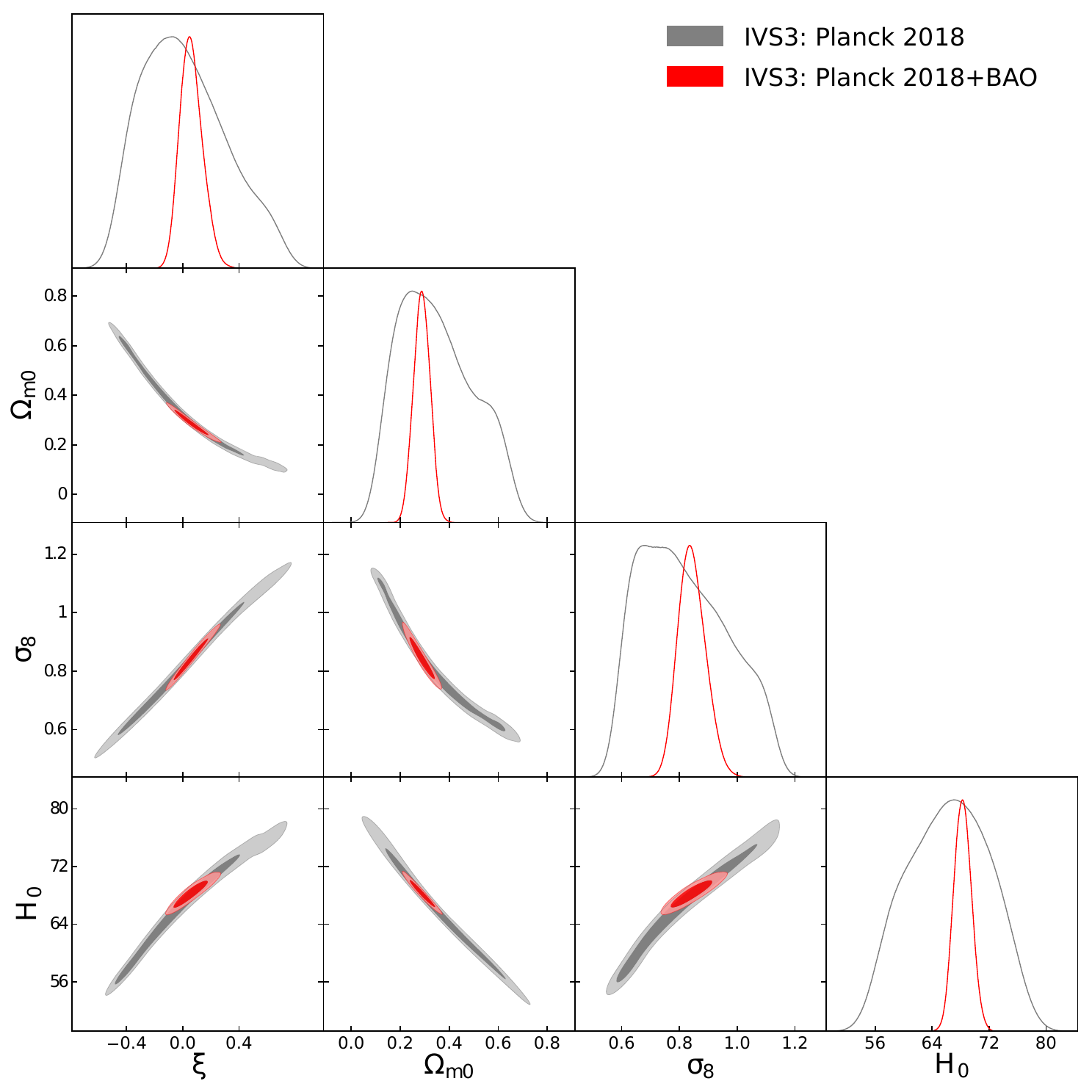}
\caption{One dimensional marginalized posterior distributions of some selective parameters and two-dimensional joint contours of various combinations of the model parameters for IVS3 have been displayed. }
\label{fig:ivs3}
\end{figure}

Finally, we consider the last interaction model in this series. Let us note that it is a nonlinear interaction model in the energy densities of the dark sectors' components unlike the previous three interaction models which are linear in the energy densities of DM and DE. So, concerning its structure, it has certain interest in this context.
In a similar way we summarize the observational constraints for this model in Table \ref{tab:IVS3} and in Fig. \ref{fig:ivs3}.

Concerning the observational constraints on the coupling
parameter $\xi$, we notice that for both Planck 2018 and Planck 2018+BAO,
$\xi =0$ is consistent within 68\% CL. For Planck 2018 alone:
$\xi =  0.012_{-    0.408}^{+    0.240}$ (68\% CL) and for
Planck 2018+BAO: $\xi =  0.062_{- 0.093}^{+    0.069} $ (68\% CL). However, both the datasets allow the
nonzero values of the coupling parameter. So, the possibility of an interaction
in the dark sector through this coupling function is equally probable.

From the constraints of $H_0$, we see that Planck 2018 alone estimates a lower
$H_0$ with very high error bars ($H_0 = 66.34_{- 6.13}^{+    6.93}$, 68\% CL), and
due to this, as one can see, within 68\% CL, it almost reaches the local estimation
of $H_0$ \cite{Riess:2019cxk}, and thus, the tension on $H_0$ is alleviated. Note that the alleviation of the tension is purely due to the error bars. However, when BAO data are added to Planck 2018, $H_0$ goes up but its error bars are reduced significantly compared to the error bars for Planck 2018, and eventually, the tension is not solved. We can say that the tension on $H_0$ is slightly weakened.

Therefore, in conclusion, within this interaction model, a mild coupling between DM and DE is supported by the observational data. Additionally, the model is also able to alleviate the $H_0$ tension due to its large error bars.

\begingroup
\squeezetable
\begin{center}
\begin{table}
\begin{tabular}{ccccccccccc}
\hline\hline
Parameters & Planck 2018 & Planck 2018+BAO \\ \hline

$\Omega_c h^2$ & $    0.1204_{-    0.0330-    0.0779}^{+    0.0565+    0.0697}$ & $    0.1109_{-    0.0100-    0.0225}^{+    0.0124+    0.0202}$  \\

$\Omega_b h^2$ & $    0.02230_{-    0.00015-    0.00031}^{+    0.00015+    0.00031}$ & $    0.02234_{-    0.00015-    0.00029}^{+    0.00015+    0.00030}$ \\

$100\theta_{MC}$ & $    1.04078_{-    0.00320-    0.00391}^{+    0.00166+    0.00470}$ & $    1.04120_{-    0.00071-    0.00118}^{+    0.00060+    0.00134}$  \\

$\tau$ & $    0.054_{-    0.0083-    0.016}^{+    0.0077+    0.016}$ & $    0.055_{-    0.0082-    0.015}^{+    0.0075+    0.017}$ \\

$n_s$ & $    0.9721_{-    0.0043-    0.0085}^{+    0.0042+    0.0084}$ & $    0.9734_{-    0.0042-    0.0080}^{+    0.0042+    0.0082}$   \\

${\rm{ln}}(10^{10} A_s)$ & $    3.055_{-    0.017-    0.032}^{+    0.016+    0.032}$ & $    3.056_{-    0.017-    0.032}^{+    0.016+    0.033}$   \\

$\xi$ & $    0.012_{-    0.408-    0.543}^{+    0.240+    0.634}$ & $    0.062_{-    0.093-    0.154}^{+    0.069+    0.165}$  \\

$\Omega_{m0}$ & $    0.349_{-    0.216-    0.265}^{+    0.118+    0.305}$ & $    0.288_{-    0.034-    0.065}^{+    0.034+    0.065}$ \\

$H_0$ & $   66.34_{-    6.13-   11.08}^{+    6.93+   10.78}$ & $   68.29_{-    1.32-    2.32}^{+    1.19+    2.46}$   \\
\hline\hline
\end{tabular}
\caption{Observational constraints at 68\% and 95\% CL on IVS3 corresponding to the interaction function $Q = 3 H \xi (\rho_c \rho_x)/(\rho_c+\rho_x)$, using the CMB data from Planck 2018 and the data from BAO.  Note that $H_0$ is in the units of Km \, s$^{-1}$ \, Mpc$^{-1}$.  }
\label{tab:IVS3}
\end{table}
\end{center}
\endgroup
\begin{figure*}
\includegraphics[width=0.45\textwidth]{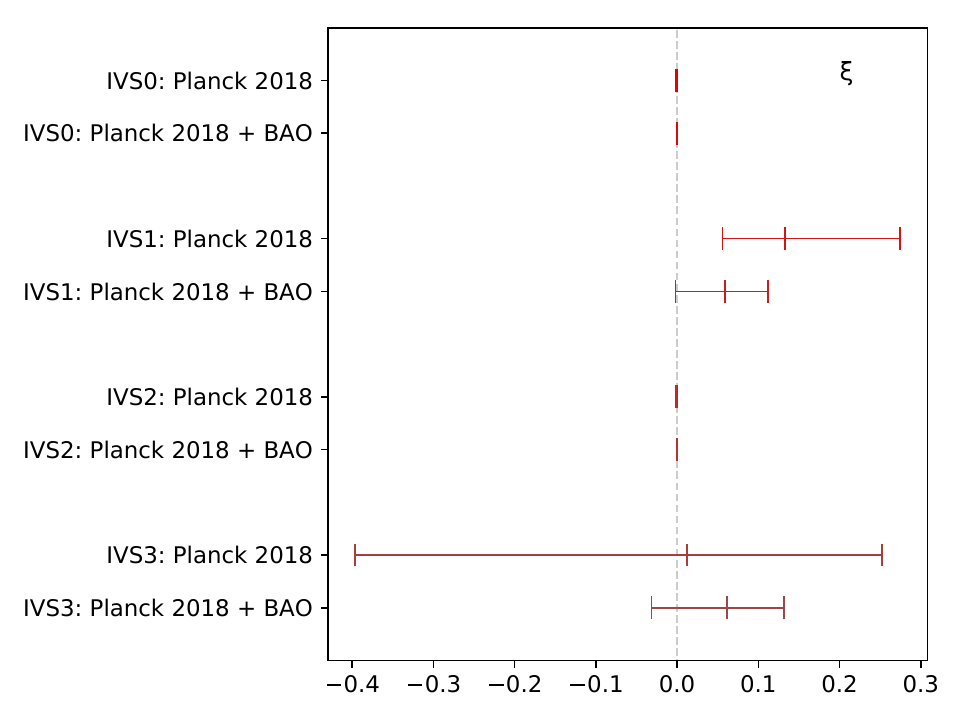}
\includegraphics[width=0.45\textwidth]{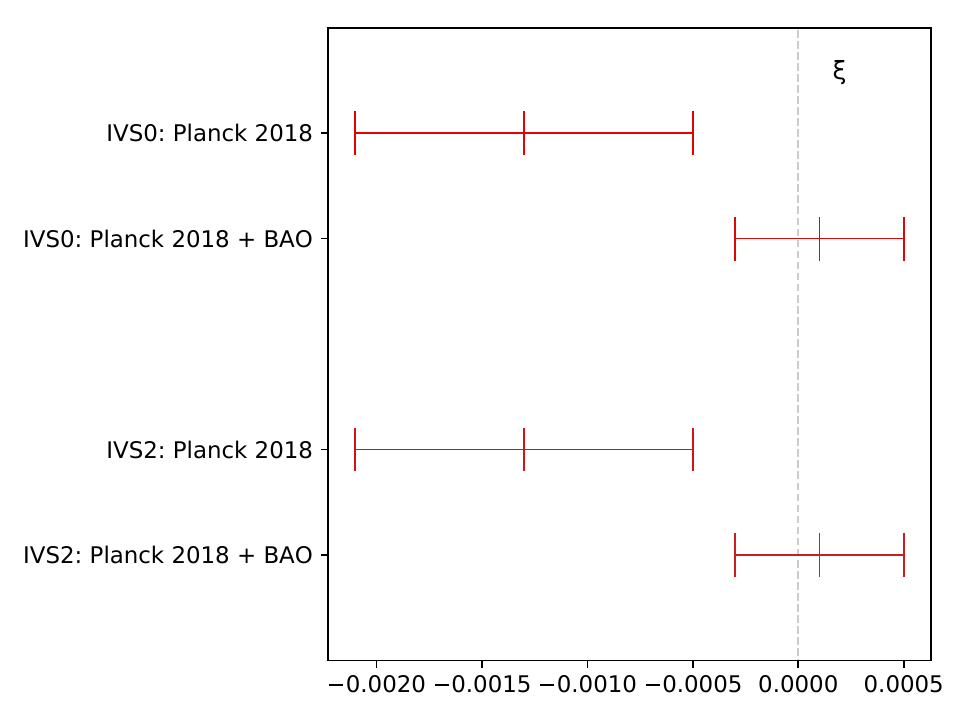}
\caption{Whisker graphs of the coupling parameter $\xi$ for all the interacting vacuum 
scenarios considering the datasets Planck 2018 and Planck 2018+BAO. The vertical dotted
line present in each graph corresponds to $\xi  = 0$. In the left graph we consider all
four models while for clarity in the right panel we have considered the whisker plot
only for the models IVS0 and IVS2. In the main text we have clarified why we show two
whisker graphs for showing the estimation of the coupling parameter, $\xi$. }
\label{fig:whiskerxi}
\end{figure*}
\begin{figure}
\includegraphics[width=0.48\textwidth]{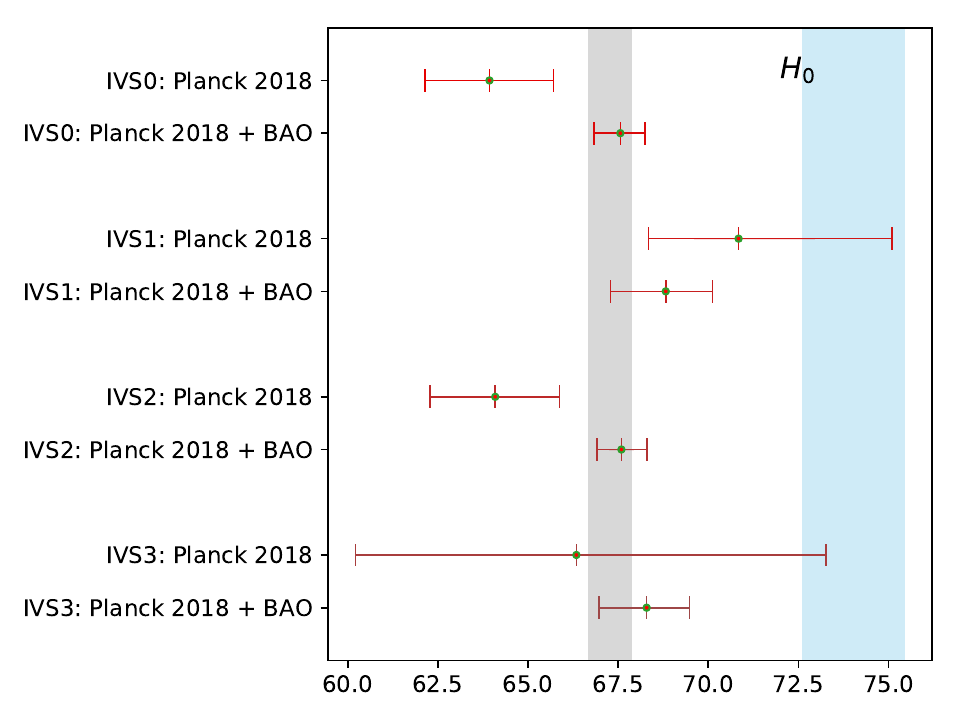}
\caption{Whisker plot showing the 68\%~CL constraints on $H_0$ for all the interacting vacuum scenarios, namely, IVS0, IVS1, IVS2 and IVS3 considering the datasets Planck 2018 and Planck 2018+BAO.
 The grey vertical band refers to estimate of $H_0$ by the Planck 2018
release~\cite{Aghanim:2018eyx} and the pale blue vertical band corresponds to the
estimation of $H_0$ (labeled as R19 in the main text) measured by SH0ES
collaboration~\cite{Riess:2019cxk}. } \label{fig:whiskerH0}
\end{figure}
\begin{figure*}
\includegraphics[width=0.34\textwidth]{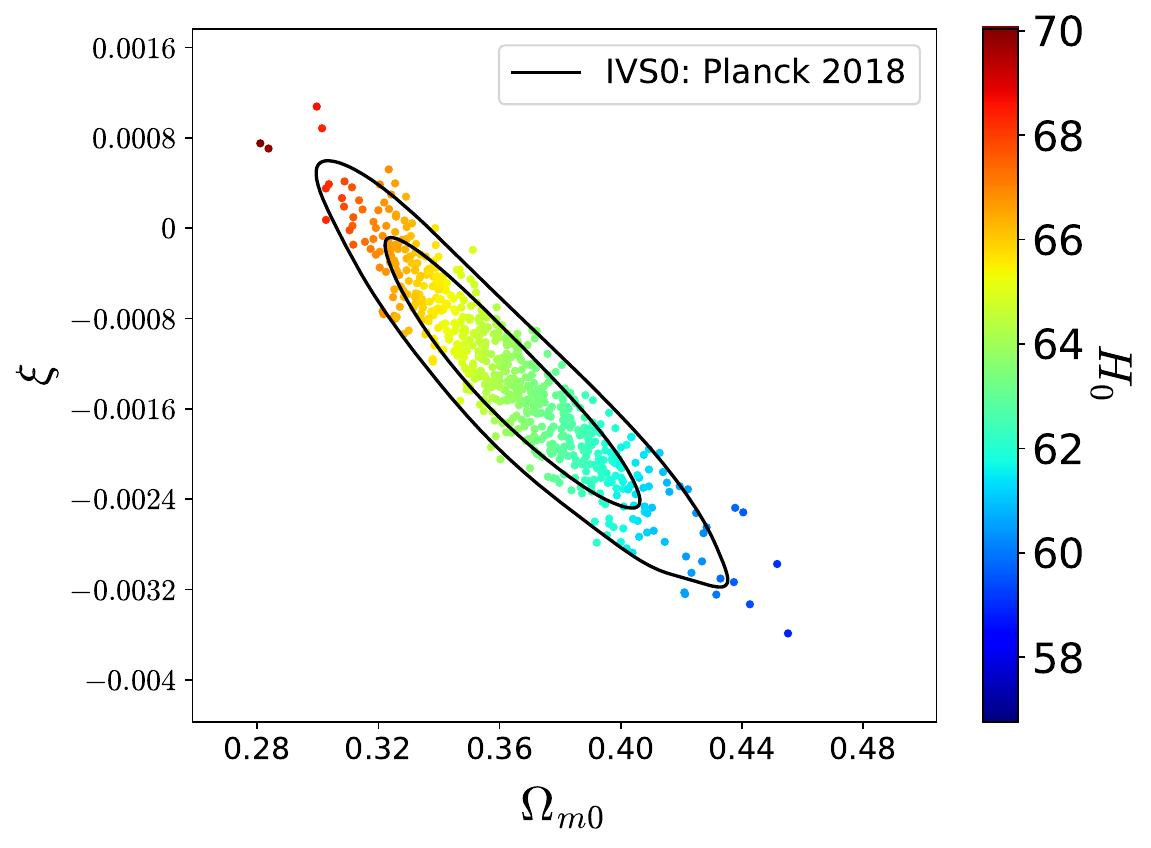}
\includegraphics[width=0.33\textwidth]{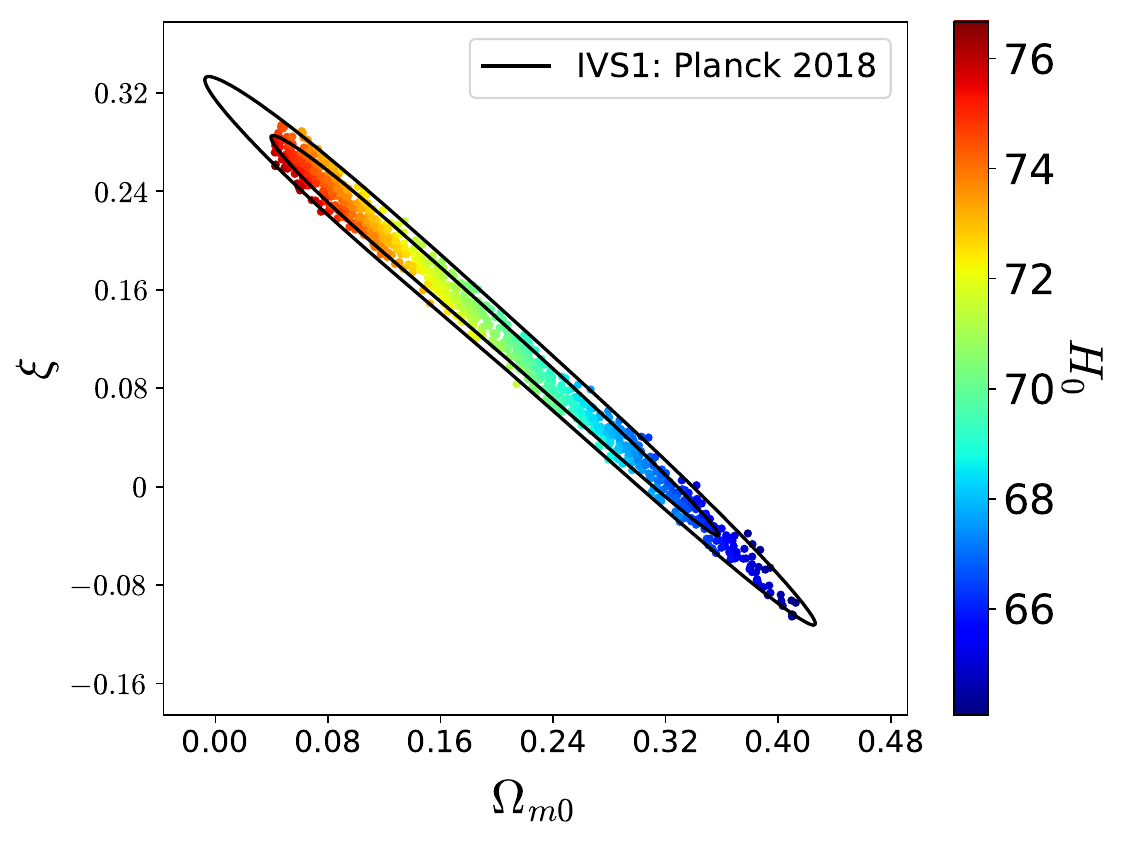}
\includegraphics[width=0.34\textwidth]{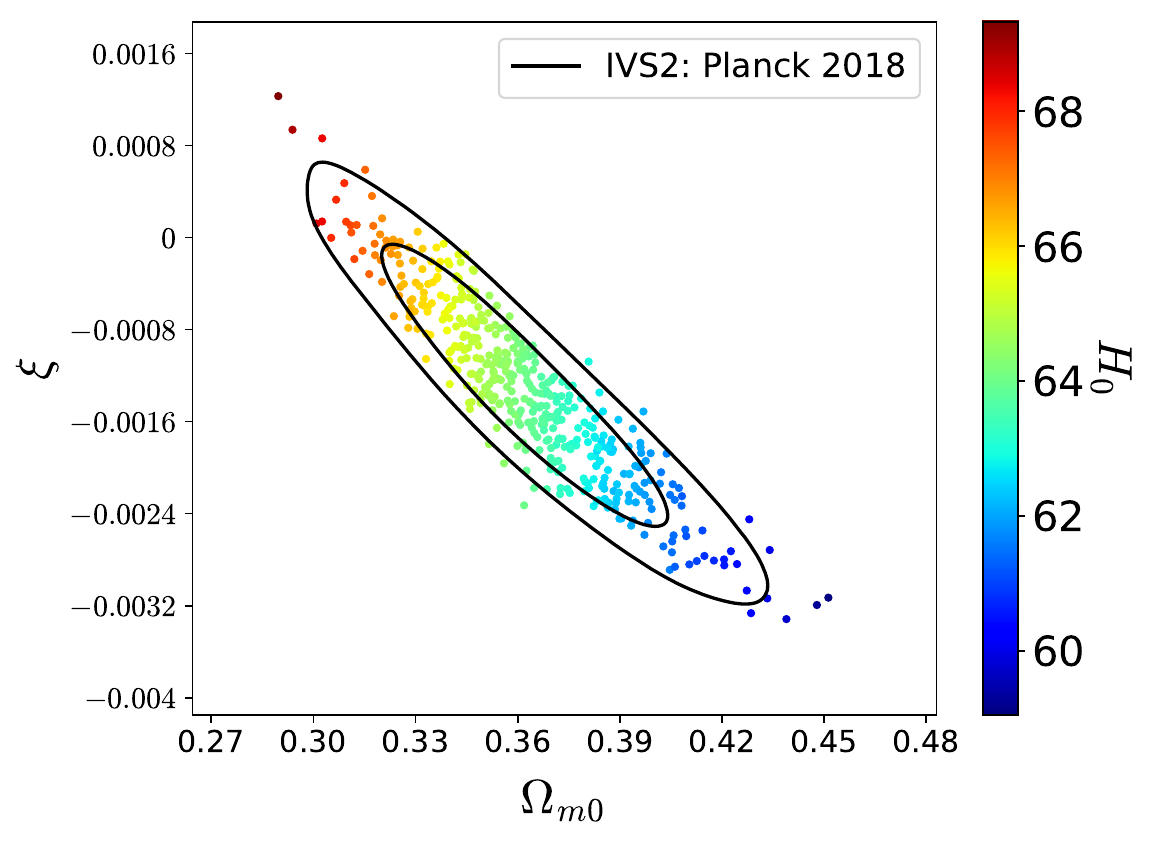}
\includegraphics[width=0.33\textwidth]{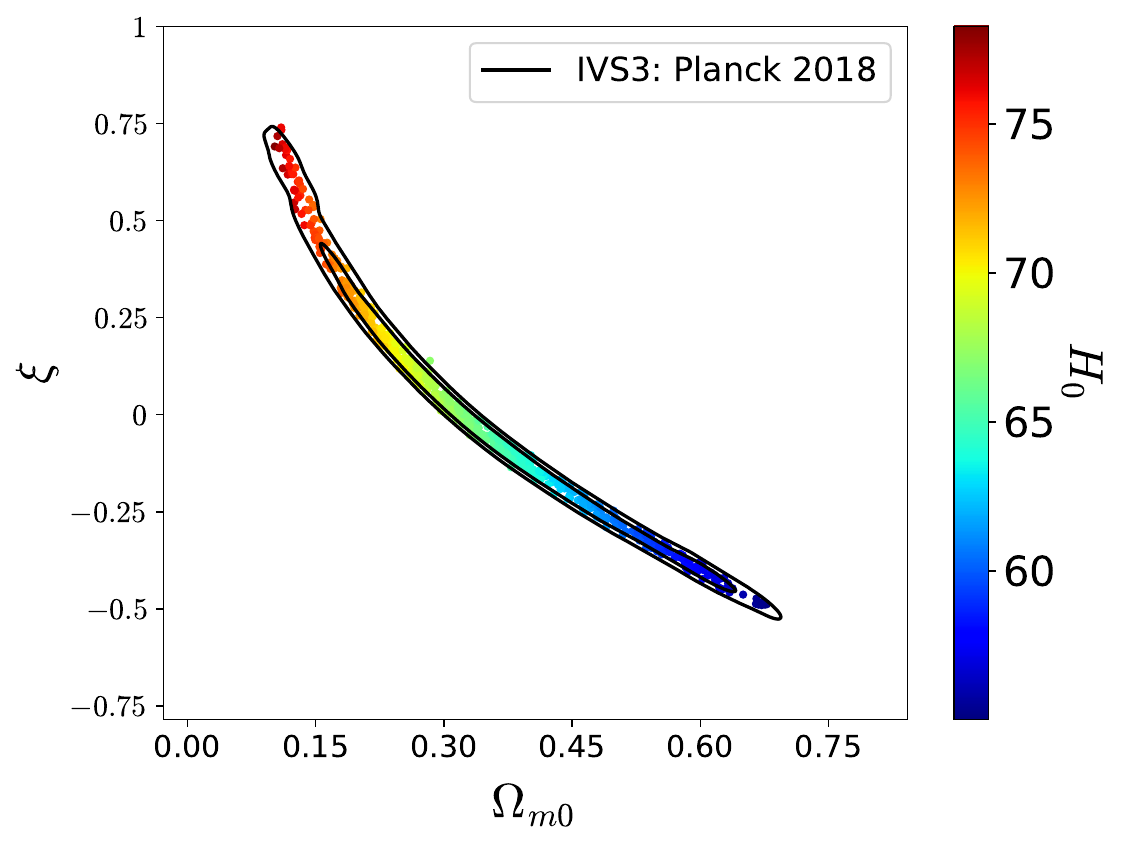}
\caption{3D scattered plots in 68\% and 95\% CL for all the IVS models in the $\xi - \Omega_{m0}$ plane colored by the $H_0$ values using the Planck 2018 data only. }
\label{fig:scattered1}
\end{figure*}
\begin{figure*}
\includegraphics[width=0.343\textwidth]{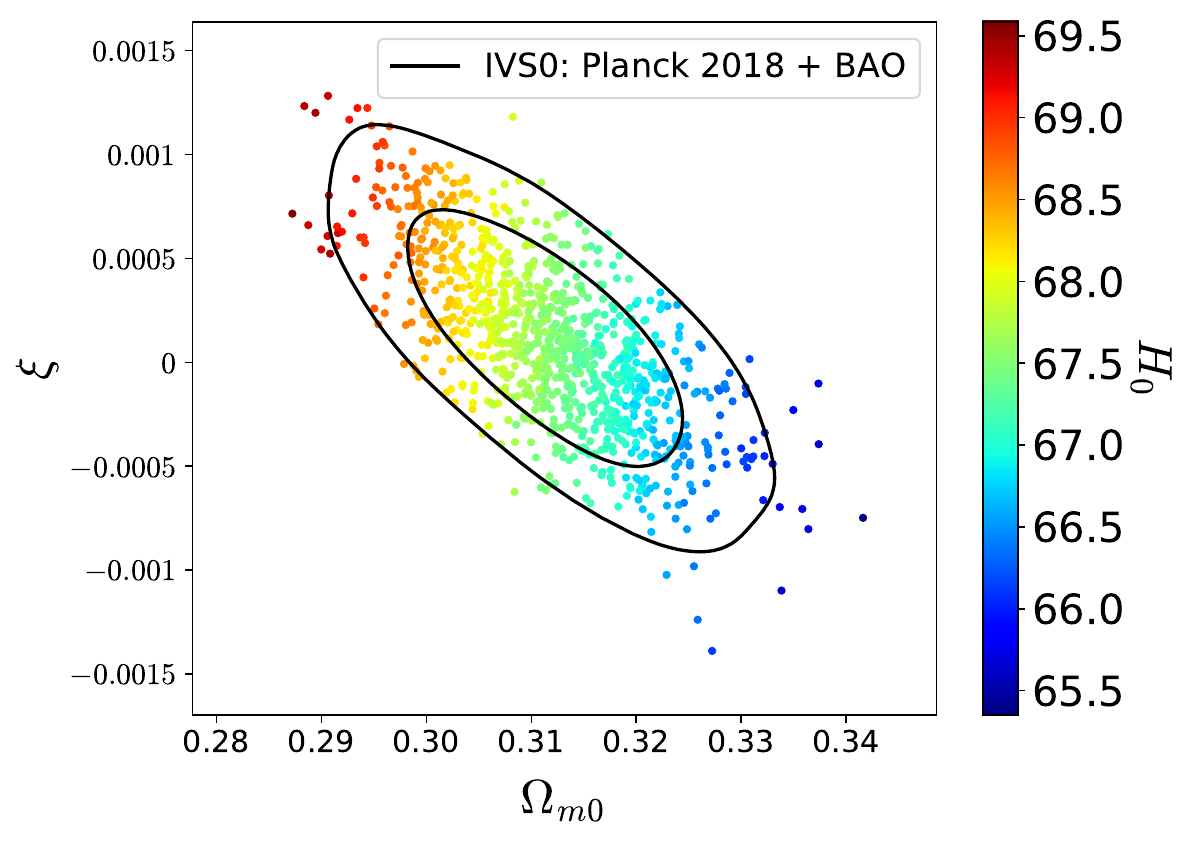}
\includegraphics[width=0.33\textwidth]{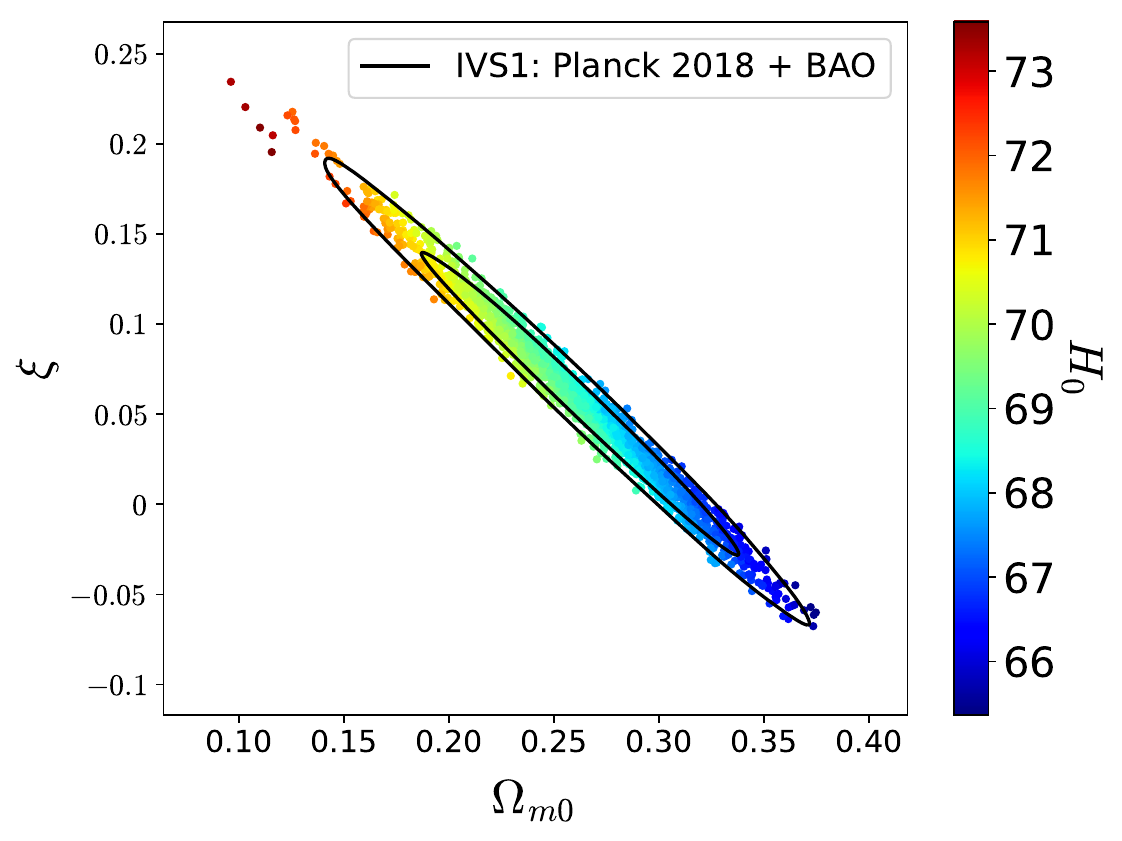}
\includegraphics[width=0.34\textwidth]{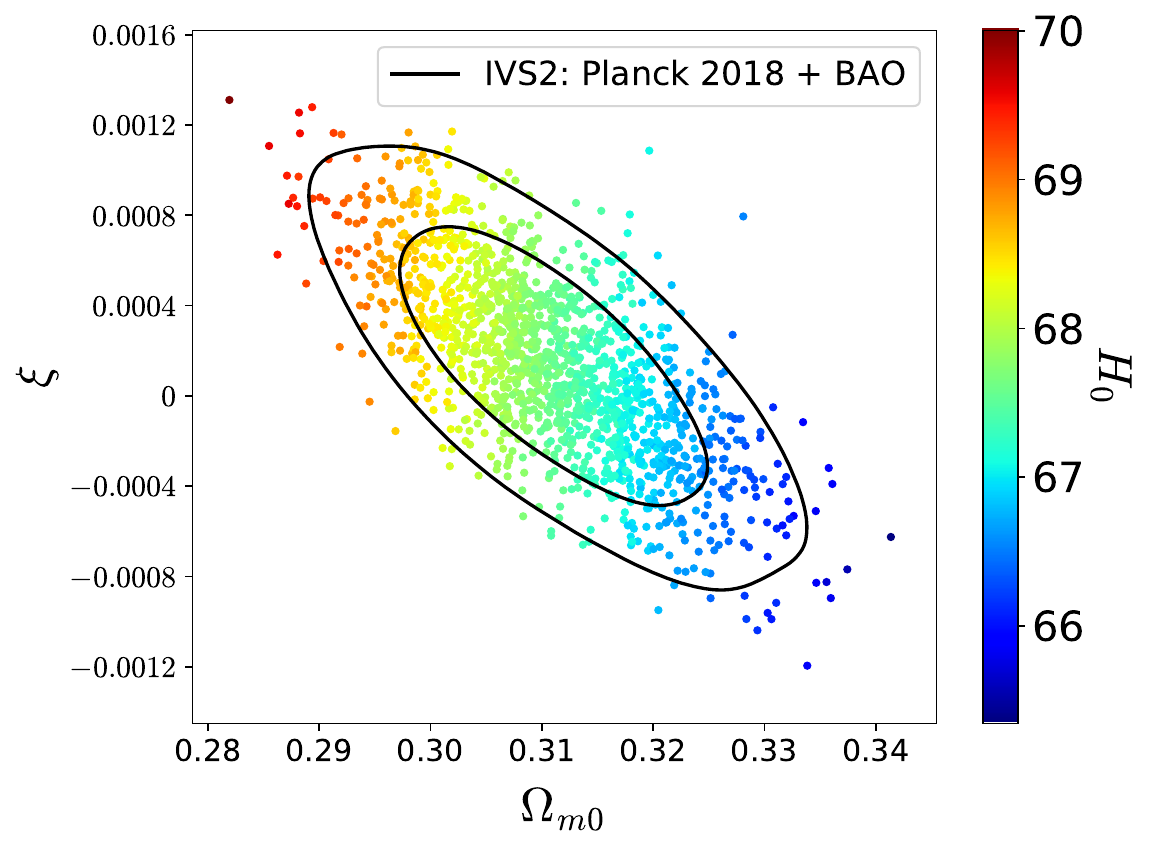}
\includegraphics[width=0.33\textwidth]{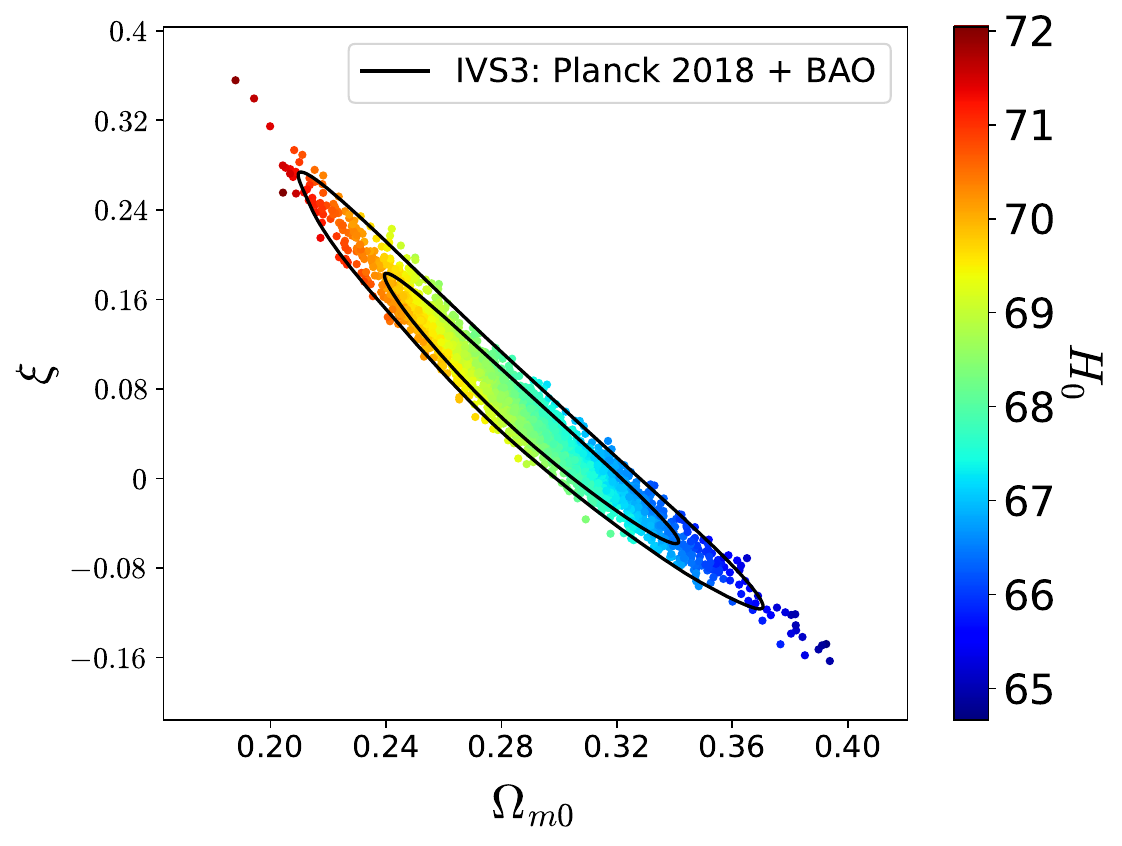}
\caption{3D scattered plots in 68\% and 95\% CL for all the IVS models in the $\xi - \Omega_{m0}$ plane colored by the $H_0$ values using the Planck 2018+BAO data only. }
\label{fig:scattered2}
\end{figure*}
\begin{figure}
\includegraphics[width=0.4\textwidth]{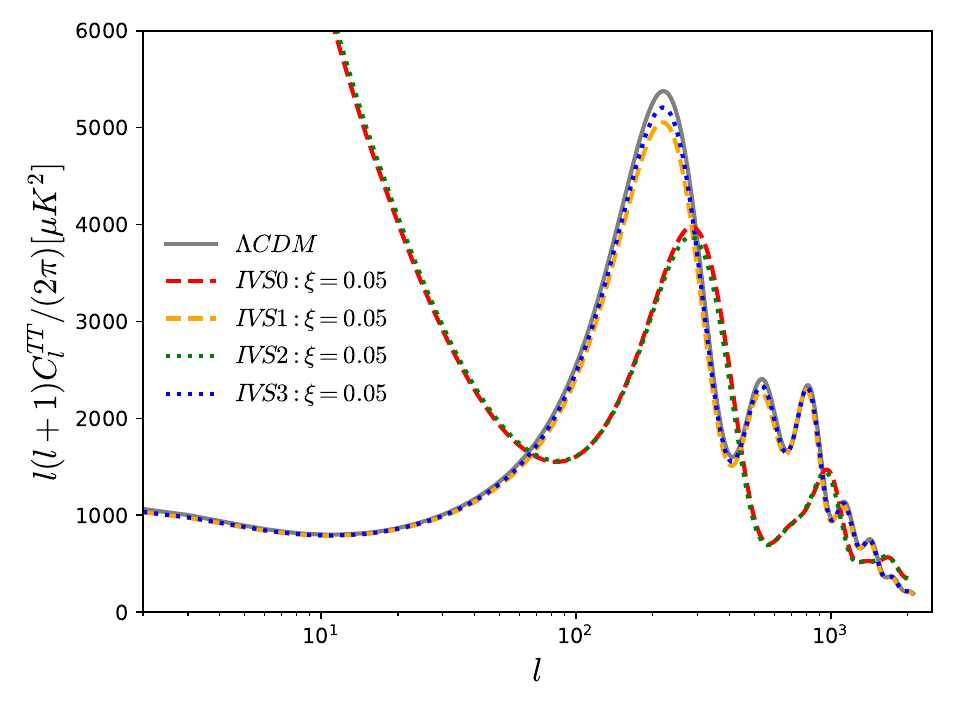}
\includegraphics[width=0.4\textwidth]{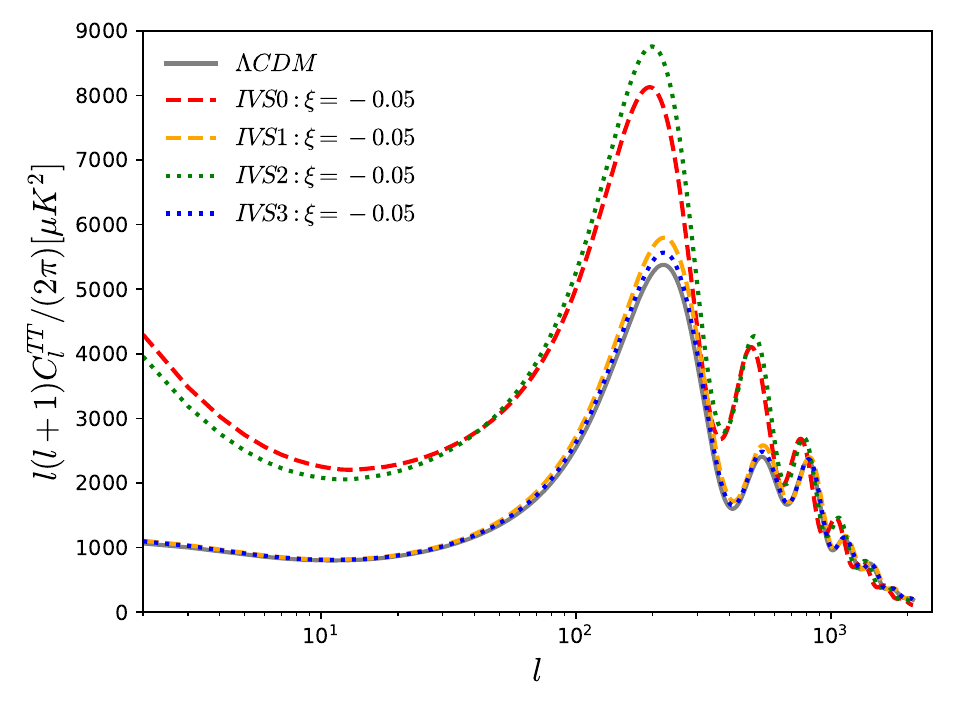}
\caption{We show the temperature anisotropy in the CMB TT spectra for different IVS models using two specific values of the dimensionless coupling parameter $\xi$. The upper panel shows the CMB TT spectra for all the models for $\xi > 0$ and the lower panel shows the same diagram but for $\xi <0$. To draw both the plots, we set the parameters values: $\Omega_{c0} = 0.28$, $\Omega_{x0} = 0.68$, $\Omega_{r0} = 0.0001$, and $\Omega_{b0} = 1- \Omega_{r0}-\Omega_{c0}-\Omega_{x0} =  0.0399$. }
\label{fig-cmbTT}
\end{figure}
\begin{figure}
\includegraphics[width=0.4\textwidth]{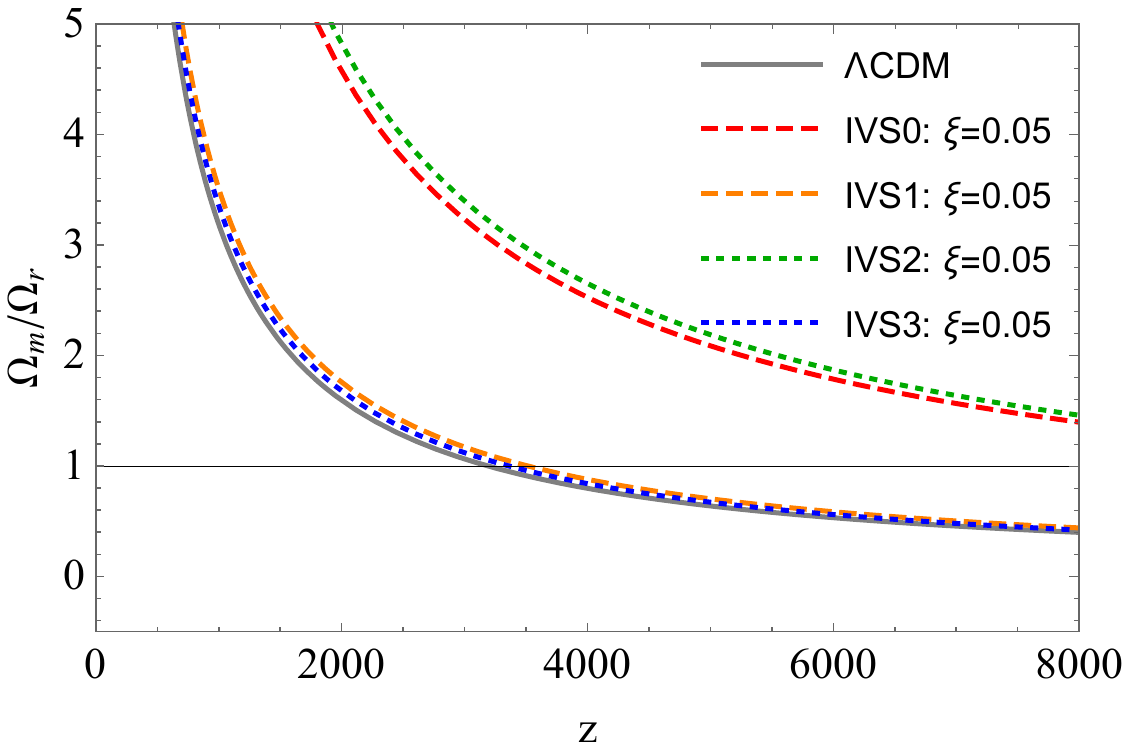}
\includegraphics[width=0.4\textwidth]{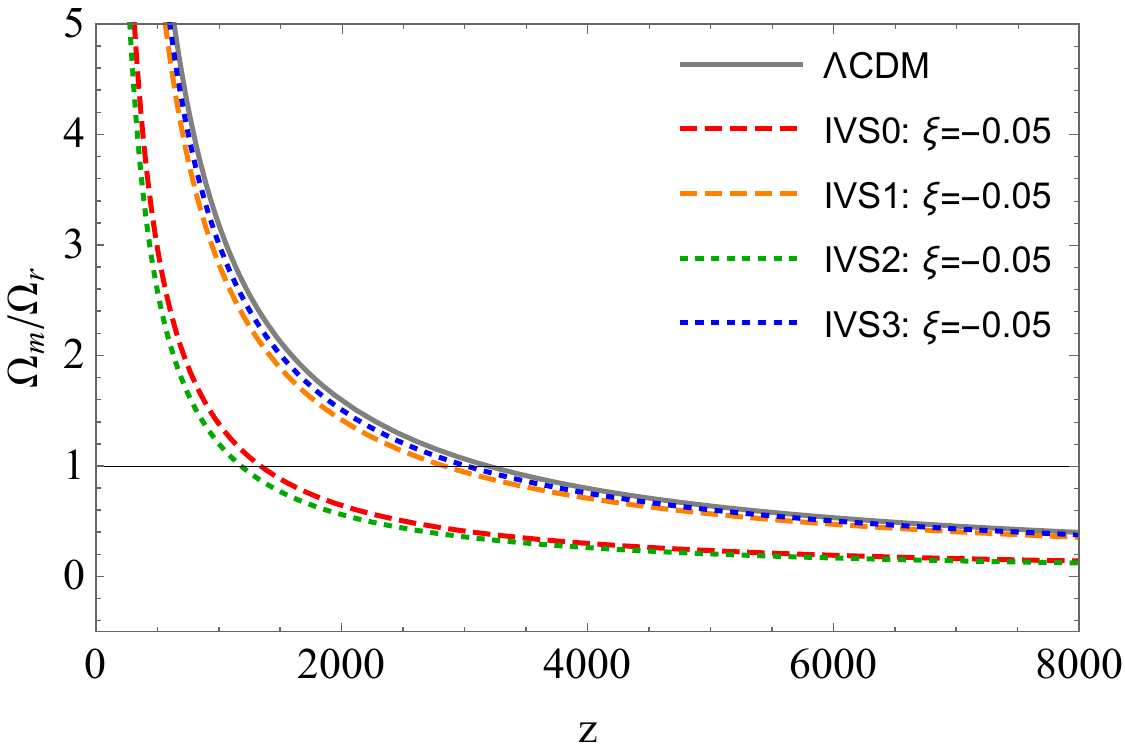}
\caption{Qualitative evolution of $\Omega_m/\Omega_r$ for various IVS scenarios  has been shown using two different values of $\xi$, namely, $\xi > 0$ (upper plot) and $\xi < 0$ (lower plot). Here, $\Omega_m$  denotes the total matter sector that means cold dark matter plus baryons, that means $\Omega_m = \Omega_c + \Omega_b$. The horizontal line denotes $\Omega_m = \Omega_r$ that means where matter density becomes equal to the radiation density. To draw both the plots, we set the following values of the parameters:  $\Omega_{c0} = 0.28$, $\Omega_{x0} = 0.68$, $\Omega_{r0} = 0.0001$, and $\Omega_{b0} = 1- \Omega_{r0}-\Omega_{c0}-\Omega_{x0} =  0.0399$. }
\label{fig-ratio}
\end{figure}
\begin{figure}
\includegraphics[width=0.4\textwidth]{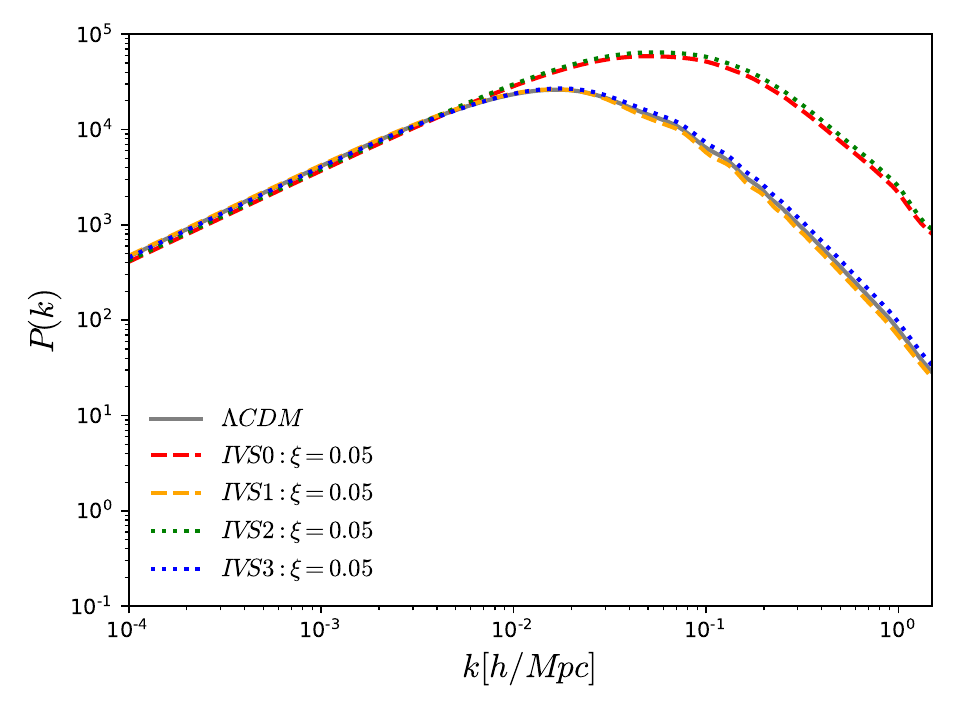}
\includegraphics[width=0.4\textwidth]{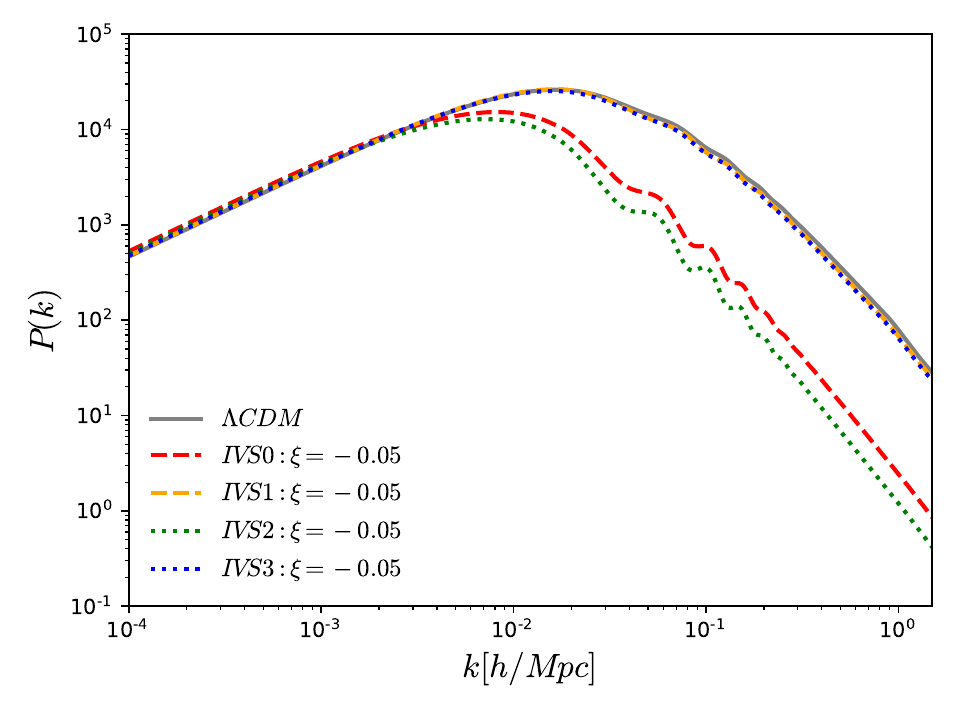}
\caption{Matter power spectrum for all the IVS scenarios using two specific values of the coupling parameter, $\xi$ has been shown. The upper plot corresponds to $\xi> 0$ while the lower plot corresponds to the case with $\xi < 0$. To draw both the plots, we set the following values of the parameters:  $\Omega_{c0} = 0.28$, $\Omega_{x0} = 0.68$, $\Omega_{r0} = 0.0001$, and $\Omega_{b0} = 1- \Omega_{r0}-\Omega_{c0}-\Omega_{x0} =  0.0399$. } 
\label{fig-matterpower}
\end{figure}

\subsection{Model comparisons}
\label{sec-model-comparison}

In the previous subsections we have described the observational constraints on the
prescribed IVS scenarios.  In this section we aim to compare the models through their
observational direction as well as we also perform a Bayesian evidence analysis in order
to test the observational viability of the models with respect to some reference model.
Since $\Lambda$CDM is the ideal choice to compare the interacting  cosmological models
under consideration, therefore, for Bayesian evidence analysis, we set $\Lambda$CDM as
the base/reference model.

In the first half of this section we compare the models focusing on their observational
constraints as well as their effects on the large scale of the universe.
In the second half of this
subsection we provide the Bayesian evidence analysis for all the models with respect to
the base $\Lambda$CDM model.

In Fig. \ref{fig:whiskerxi},  we show the whisker graphs for the coupling parameter $\xi
$ of all the interacting scenarios, namely, IVS0, IVS1, IVS2 and IVS3, considering the
analyses Planck 2018 and Planck 2018+BAO. Let us note that for IVS0 and IVS2, the region
of $\xi$ coincides with the vertical line representing $\xi = 0$. The reason for such an
overlap is that, for both IVS0 and IVS2, the constraints on $\xi$ are very close to zero
(see Tables \ref{tab:IVS0} and \ref{tab:IVS2} for this purpose). That is why in the
right graph of  Fig. \ref{fig:whiskerxi} we have separately shown the whisker plot for
IVS0 and IVS2. Now, from the left graph, one can safely conclude that models IVS1 and
IVS3 assume similar constraints, although for IVS3, $\xi < 0$ is allowed. While the
constraints from IVS0 and IVS2 are almost same. This is pretty clear from the right
graph of Fig. \ref{fig:whiskerxi}.

In order to understand how the present IVS models could be effective in
alleviating/solving the $H_0$ tension,  in Fig. \ref{fig:whiskerH0}, we show the whisker
graph for $H_0$ (at 68\% CL) for Planck 2018 and Planck 2018+BAO.
 The grey vertical band corresponds to  $H_0$ estimated by the Planck 2018
 release~\cite{Aghanim:2018eyx} and the pale blue vertical band
 denotes the $H_0$ estimation by SH0ES collaboration~\cite{Riess:2019cxk}. Let us first consider the response of all interacting scenarios to the $H_0$ tension for Planck 2018 data only. Looking at the estimated values of $H_0$ from different interacting scenarios, namely, $H_0 =   63.93_{-1.79}^{+    1.78}$ Km \, s$^{-1}$ \, Mpc$^{-1}$  at 68\% CL (for IVS0); $H_0 = 70.84^{+4.26}_{-2.50}$ Km \, s$^{-1}$ \, Mpc$^{-1}$ at 68\% CL (for IVS1); $H_0 = 64.09_{- 1.81}^{+1.78} $ Km \, s$^{-1}$ \, $Mpc^{-1}$ at 68\% CL (for IVS2); and $H_0 = 66.34_{-11.08}^{+ 6.93}$   Km \, s$^{-1}$ \, Mpc$^{-1}$ at 68\% CL (for IVS3), one can clearly see that for IVS1, the mean value of $H_0$ is very high compared to other interacting scenarios in this work and the minimal $\Lambda$CDM cosmology by Planck (where $H_0= 67.27 \pm 0.60$ Km \, s$^{-1}$ \, Mpc$^{-1}$ at 68\% CL for Planck TT,TE,EE+lowE \cite{Aghanim:2018eyx}). Now, due to the large error bars on $H_0$ for IVS1, the tension on $H_0$ is much alleviated. This is pretty clear from the whisker graph shown in Fig. \ref{fig:whiskerH0}. However, for IVS3, although $H_0$ has very high error bars ($H_0 = 66.34_{-11.08}^{+ 6.93}$   Km \, s$^{-1}$ \, Mpc$^{-1}$ at 68\% CL), but due to very small mean value ($H_0 \sim 66 $   Km \, s$^{-1}$ \, Mpc$^{-1}$), one cannot strongly argue that the tension on $H_0$ is satisfactorily alleviated. One can actually say that the tension on $H_0$ in this case is very mildly weakened just because of such high error bars on $H_0$. When BAO data are added to Planck 2018, the combined dataset Planck 2018+BAO, as one can clearly see from the whisker graph in Fig. \ref{fig:whiskerH0}, cannot alleviate the tension on $H_0$ for IVS0 and IVS2. However, for IVS1 and IVS3, one can say that the tension on $H_0$ is very mildly weakened (see Fig.  \ref{fig:whiskerH0}) due to slightly higher mean values of $H_0$ together with its higher error bars (see the Planck 2018+BAO columns in Tables \ref{tab:IVS1} and \ref{tab:IVS3}) compared to the Planck's estimation within the minimal $\Lambda$CDM cosmology \cite{Aghanim:2018eyx}.   
 Finally, one might be interested to look at
the 3D scattered plots for the IVS models shown in Fig. \ref{fig:scattered1} and
\ref{fig:scattered2}. From the scattered plots displayed in Fig. \ref{fig:scattered1}
and \ref{fig:scattered2} one can understand the behaviour of the coupling parameter,
$\xi$, with higher and lower values of the Hubble constant, $H_0$. For both Planck 2018
and Planck 2018+BAO, as we can see, for higher values of $H_0$, $\xi$ assumes (although
mildly) positive values, indicating an energy transfer from pressureless DM to DE. For
lower values of $H_0$, exactly opposite scenario is confirmed. The scattered plots
actually give a nice statistical comparisons between the models. 

We also display  in Fig. \ref{fig-cmbTT} the temperature anisotropy in the CMB TT
spectra for all the IVS models using some specific values of the dimensionless coupling
parameter $\xi$. In the upper panel of  Fig. \ref{fig-cmbTT} we show the CMB TT spectra
for a specific positive value of $\xi = 0.05$   and in the lower plot of Fig.
\ref{fig-cmbTT} we display the same physical quantity for negative value of the coupling
parameter, that means, $\xi = - 0.05$.
 For comparison purpose,
 we have also included the non-interacting $\Lambda$CDM scenario
($\xi =0$). One can quickly realize that IVS0 and IVS2 are quite different compared to
IVS1 and IVS3. Let us describe the physics behind the plots. From the upper plot of Fig.
\ref{fig-cmbTT}, one can see that for all IVS models, the heights of the first acoustic
peak in the CMB TT spectrum, are less than the height of the first acoustic peak for the
non-interacting $\Lambda$CDM model. One can understand this phenomena from the evolution
of the matter-radiation equality for all the IVS models.
 It is clear that if we add an interaction in the dark sector, the evolution of
 the CDM sector
  will not follow its usual
evolution which is $\rho_c \propto a^{-3}$, hence, the evolution of the matter sector,
$\Omega_m \; (= \Omega_c +\Omega_b)$ that includes CDM and baryons, will definitely
change from its usual evolution, and hence the matter-radiation equality will alter. If
one looks at the upper plot of Fig. \ref{fig-ratio}, it is clear that for all IVS
models, the matter-radiation equality happens earlier for $\xi >0$ compared to the
non-interacting $\Lambda$CDM model. Due to earlier matter-radiation equality, the sound
horizon is decreased, hence, for the present IVS models, the first peak in the CMB TT
spectrum is decreased. For IVS0 and IVS2, the matter-radiation equality happens much
earlier compared to IVS1 and IVS3, and this has been encoded in the CMB TT spectrum in
terms of significant reduction of the first peak compared to other two IVS models. On
the other hand, for $\xi < 0$, exactly the opposite scenario happens in the CMB TT
spectrum (see the lower panel of  Fig. \ref{fig-cmbTT}) and this behaviour become clear
when one looks at the corresponding matter-radiation equality presented in the lower
plot of Fig. \ref{fig-ratio}. 

We also investigate the effects of the IVS models in the matter power spectrum. In
Fig. \ref{fig-matterpower} we show the matter power spectrum for all the IVS models for
two specific values of the coupling parameter, namely, $\xi >0$ (upper panel of Fig.
\ref{fig-matterpower}) and $\xi <0$ (lower panel of Fig. \ref{fig-matterpower}). We
again find that the behaviour of IVS0 and IVS2 are completely different (in fact,
violent) compared to the other IVS models. To understand the behaviour of various IVS
models compared to the no-interaction scenario, we have considered the matter power
spectrum for the  non-interacting $\Lambda$CDM model. From the upper plot of Fig.
\ref{fig-matterpower} we see that the amplitude of the matter power spectrum for all IVS
models increases compared to the $\xi  =0$ case. The significant increase in the matter
power spectrum
 is transparent
 for IVS0 and IVS2 while for other two IVS models, it is
quite difficult to understand the changes in the matter power spectrum from the
non-interacting scenario ($\xi =0$). The enhancement in the matter power spectrum is for
the earlier matter-radiation equality, see the upper plot of Fig. \ref{fig-ratio}. The
reverse situation occurs for $\xi < 0$ (see the lower plot of Fig.
\ref{fig-matterpower}). In this case the matter power spectrum are suppressed and this
again corresponds to the late matter-radiation equality as shown in the lower plot of
Fig. \ref{fig-ratio}.

Thus, from the behaviour of the IVS models presented in the CMB TT and matter power
spectra shown respectively in Fig. \ref{fig-cmbTT} and \ref{fig-matterpower}, it is
clearly pronounced that models IVS0 and IVS2 are significantly different from the rest
two IVS models, namely, IVS1 and IVS3, and additionally, they are very far from the
non-interacting $\Lambda$CDM model which is only detected through the analysis of
formation of structure of the universe.

Finally, we perform the Bayesian evidence analysis for a better understanding on the
models with respect to some reference model. To calculate the evidences we use the
MCEvidence \cite{Heavens:2017afc,Heavens:2017hkr}, a cosmological code for computating
the evidences of the interacting scenarios (also see \cite{Pan:2017zoh,Yang:2018qmz} for
detailed descriptions). To quantify the observational support of the models, we use the
revised Jeffrey's scale through different values of $\ln B_{ij}$. The strength of
evidence of the underlying model ($M_j$) with respect to the reference  $\Lambda$CDM
scenario ($M_i$) is characterized as follows \cite{Kass:1995loi}: (i) for $0 \leq \ln
B_{ij} < 1$, a weak evidence, (ii) for $1 \leq \ln B_{ij} < 3$, a Definite/Positive
evidence; (iii) for $3 \leq \ln B_{ij} < 5$, a strong evidence, and  (iv) for $\ln
B_{ij} \geq 5$, a very strong evidence for the reference $\Lambda$CDM model (``$i$'')
against the underlying model (here the interacting scenario) is considered.  In Table
\ref{tab:bayesian} we have summarized the values of $\ln B_{ij}$. From Table
\ref{tab:bayesian} we find that $\Lambda$CDM is favored
 by the observational data over
all the IVS models, but the models, namely,  IVS1 and IVS3 are relatively close to
$\Lambda$CDM compared to the remaining two IVS models (IVS0 and IVS2). So, in summary, the models IVS1 and IVS3 have some importance in the literature in light of the Bayesian evidence analysis.

\begin{table}
\begin{center}
\begin{tabular}{ccccccccc}                                      \hline\hline
Dataset & Model &~~ $\ln B_{ij}$ & \\
\hline

Planck 2018 & IVS0 & $4.2$ \\
Planck 2018+BAO & IVS0 & $6.7$ \\

\hline\hline

Planck 2018 & IVS1 & $1.0$ \\
Planck 2018+BAO & IVS1 & $1.8$ \\

\hline\hline

Planck 2018 & IVS2 & $2.8$  \\
Planck 2018+BAO & IVS2 & $6.7$  \\

\hline \hline

Planck 2018 & IVS3 & $0.3$  \\
Planck 2018+BAO & IVS3 & $1.4$  \\

\hline \hline

\end{tabular}
\caption{Summary of $\ln B_{ij}$ values computed for the $\Lambda$CDM model with respect to IVS0, IVS1, IVS2 and IVS3.}
\label{tab:bayesian}
\end{center}
\end{table}

\section{Concluding remarks}
\label{sec-summary}

Interacting DM -- DE models have gained potential interest for explaining various
cosmological puzzles beginning from the  cosmic coincidence problem to the $H_0$
tension. It has been almost 20 years as of now, interacting models have been
investigated by various investigators. The interacting models are entirely dependent on
the interaction function, $Q$, that determines the rate of energy transfer between the
dark sectors DM and DE. Despite of a lot of investigations in this context, a
fundamental question -- what should be the possible functional form of $Q$ -- is still
unknown to the cosmological community. Since the nature of DM and DE are unknown, it is
very difficult to extract the exact functional form for the interaction function. Thus,
the easiest approach followed from earlier to present time, is to assume some
phenomenological functions for $Q$ and then to test them using the available
cosmological data.  The lack of a definite mechanism to construct the interaction
functions raises questions over the interaction models. This motivated us to investigate
the interaction models from the field theoretical arguments with an aim to search for a
valid route to find out the models that are widely used. Our answer is affirmative in
this direction. We have shown that
various linear and nonlinear interaction functions that have been widely examined in the
past and present, can be derived. {\it This is the main essence of this work and
probably this is the first time in the literature where we show the exact derivations of
some very well known interaction models having a solid field theoretic ground.}

We then examine the interaction models using CMB data from Planck 2018 final release
and with the BAO data. The inclusion of BAO to CMB is motivated to break the degeneracies in some parameters that may exist during the analysis with
CMB data alone.
The results are summarized in Tables \ref{tab:IVS0},
\ref{tab:IVS1}, \ref{tab:IVS2} and \ref{tab:IVS3}.   Our analyses show that although both Planck 2018 and Planck 2018+BAO mildly allow a non-zero interaction in the dark sector but $\xi = 0$ seems to be the most consistent picture.
We also find that the second interaction model, namely, IVS1 is the most
promising candidate to alleviate the $H_0$ tension in an effective way.
The models have been investigated further through their effects on the CMB TT and matter power spectrum. Such an analysis is really important because this offers more insights on the models. Our analyses clearly depict that IVS0 and IVS2 are different compared to other models. We found that presence of an interaction in the dark sector alters the matter-radiation equality and hence this effects are encoded in the CMB TT and matter power spectrum. We  notice that IVS1 and IVS3 are relatively close to the $\Lambda$CDM model.

Finally, we perform a qualitative comparisons between the IVS models through the
observational constraints and the Bayesian evidence analysis with respect to the
reference $\Lambda$CDM scenario. We find that IVS0 and IVS2 behave similarly, on the
other hand, IVS1 and IVS3 behave similarly, but the last two models have essential
advantages when we make the Bayesian evidence analysis. However, it is true that the
$\Lambda$CDM scenario is still preferred over the IVS models.

\section{Acknowledgments}
The authors thank the referee for some useful comments that helped us to improve the quality of the manuscript. SP has been supported by the Mathematical Research Impact-Centric Support Scheme
(MATRICS), File No. MTR/2018/000940, given by the Science and Engineering Research Board
(SERB), Govt. of India. WY was  supported by the
National Natural Science Foundation of China under Grants
No. 11705079 and No. 11647153.


\end{document}